\documentclass[aps,superscriptaddress,nofootinbib]{revtex4}

\usepackage{multirow}
\usepackage[T1]{fontenc}
\usepackage[utf8]{inputenc}
\usepackage{mathtools}
\usepackage{graphicx}
\usepackage{subcaption}
\usepackage[dvipsnames]{xcolor}
\usepackage{float}
\usepackage{soul}
\usepackage{amsmath,amssymb}
\usepackage{comment}
\usepackage[colorlinks=true, pdfstartview=FitV, linkcolor=blue, citecolor=blue, urlcolor=magenta, breaklinks=true]{hyperref}
\usepackage{orcidlink}
\usepackage{pifont}
\usepackage{bm}

\begin{document}

\title{Constraining the Generalized Tolman-Oppenheimer-Volkoff (GTOV) equation with Bayesian analysis}

\author{Franciele M. da Silva \orcidlink{0000-0003-2568-2901}} 
\email{franmdasilva@gmail.com}

\affiliation{Departamento de F\'isica, CFM - Universidade Federal de Santa Catarina; \\ C.P. 476, CEP 88.040-900, Florian\'opolis, SC, Brazil.}

\author{F\'abio K\"opp   \orcidlink{0000-0001-9970-4339}                                       }
\email{fabiokopp@proton.me}

\affiliation{Departamento de F\'isica, CFM - Universidade Federal de Santa Catarina; \\ C.P. 476, CEP 88.040-900, Florian\'opolis, SC, Brazil.}

\author{Marcelo D. Alloy}
\email{marcelo.alloy@ufsc.br}

\affiliation{Departamento de F\'isica, CFM - Universidade Federal de Santa Catarina; \\ C.P. 476, CEP 88.040-900, Florian\'opolis, SC, Brazil.}

\author{Luis C. N. Santos \orcidlink{0000-0002-6129-1820}}
\email{luis.santos@ufsc.br}

\affiliation{Departamento de F\'isica, CFM - Universidade Federal de Santa Catarina; \\ C.P. 476, CEP 88.040-900, Florian\'opolis, SC, Brazil.}

\author{Adamu Issifu \orcidlink{0000-0002-2843-835X}} 
\email{ai@academico.ufpb.br}

\affiliation{Departamento de F\'isica e Laborat\'orio de Computa\c c\~ao Cient\'ifica Avan\c cada e Modelamento (Lab-CCAM), Instituto Tecnol\'ogico de Aeron\'autica, DCTA, 12228-900, S\~ao Jos\'e dos Campos, SP, Brazil} 

\author{Clésio E. Mota \orcidlink{0000-0002-8616-0894}}
\email{clesio200915@hotmail.com}

\affiliation{Departamento de F\'isica, CFM - Universidade Federal de Santa Catarina; \\ C.P. 476, CEP 88.040-900, Florian\'opolis, SC, Brazil.}

\author{D\'ebora P. Menezes \orcidlink{0000-0003-0730-6689}}
\email{debora.p.m@ufsc.br}

\affiliation{Departamento de F\'isica, CFM - Universidade Federal de Santa Catarina; \\ C.P. 476, CEP 88.040-900, Florian\'opolis, SC, Brazil.}

\begin{abstract}
In this work, we constrain the values of the parameters of the Generalized Tolman-Oppenheimer-Volkoff (GTOV) equation through Bayesian inference. We use the mass and radius data from the Neutron Star Interior Composition Explorer (NICER) for PSR J0740$+$6620 and PSR J0030$+$0451, as well as the mass, radius, and dimensionless tidal deformability from the gravitational wave (GW) events GW190814 and GW170817. We use two distinct parameterizations of the extended non-linear Walecka model (eNLW) with and without hyperons. The GTOV employed for the study contains additional free parameters with different physical motivations. Two possible scenarios are considered in our analysis: conservative and speculative. In the first case, we take into account the most reliable neutron star (NS) data from NICER and the GW170817 event. In the second case, we consider the possibility that the compact object with a mass of  $2.54 M_{\odot}$ in the GW190814 event is an NS. Our findings show significant improvements in the physical quantities analyzed, leading to better agreement with the observational data compared to the results obtained using the TOV equation. 
\end{abstract}

\maketitle

\section{Introduction} \label{introduction}

The study of NSs has experienced some important progress in recent years. A significant positive impact was achieved after the implementation of NICER~\cite{NICER} on the International Space Station (ISS). One of the open problems in astrophysics is constructing a multipurpose equation of state (EoS)~\cite{CompOSECoreTeam:2022ddl} based on our knowledge of microphysics to serve as a foundation for applications in astrophysical simulations. To construct an EoS that describes the current data and future observational advancements of compact objects requires an understanding of strongly interacting matter at high baryon densities governed by the quantum chromodynamics (QCD) theory which is yet to be fully understood. Nonetheless, the unset of data from NICER has contributed significantly to constraining the EoSs, since it provides simultaneous estimates for the mass and radius of the pulsars PSR J0030$+$0451~\cite{riley2019nicer} and PSR J0740$+$6620~\cite{riley2021nicer}. This has formed the basis for reducing the EoS parameterizations in the literature~\cite{raaijmakers2019nicer,bogdanov2019constraining,miller2019psr} to a larger extent. Another important advancement in NS investigation is the detection of gravitational waves (GWs) from the merger of binary NSs. The main result is the GW detection in the GW170817~\cite{LIGOScientific:2017vwq} event, which has been confirmed by multiple observatories in several wavelengths of the electromagnetic spectrum~\cite{abbott2017gravitational,abbott2017multi}. In addition to imposing more constraints on the mass and radius of the NSs, the GW detection led to the first measurement of the tidal deformability of the NSs, further restricting the EoS~\cite{LIGOScientific:2018cki,lourencco2019consistent,lourencco2020consistent,lopes2021hyperonic,raaijmakers2020constraining,raaijmakers2021constraints,radice2018gw170817,malik2018gw170817,tews2018critical}. Aside the binary NS merger event, the detection of merger events with at least one NS participating have also been reported in events GW190425~\cite{LIGOScientific:2020aai}, GW200105 and GW200115~\cite{abbott2021observation}. 

Due to the high densities found in NSs, it is necessary to use the theory of General Relativity (GR) to appropriately describe these objects. GR is one of the most successful classical theories in history~\cite{GR_tests}, however, there is evidence that it may not be complete. Modifications  arise mainly from Dark Energy~\cite{copeland2006dynamics,dark2016dark}, Dark Matter~\cite{bertone2018history}, Inflation~\cite{guth2000inflation}, Large Scale Structure~\cite{springel2006large}, and an effective description of Quantum Gravity~\cite{carlip2001quantum,addazi2022quantum}. These modifications are known as extended, modified, or alternative theories of gravity ~\cite{Capozziello:2011et,clifton2012modified,sotiriou2010f,olmo2020stellar,shankaranarayanan2022modified}. To date, several of these theories have been developed and tested in varied physical settings. In special, tests of these modified theories on NSs are of great interest due to the strong gravitational field of these objects~\cite{olmo2020stellar,Silva2014,Folomeev2018,paper2,paper1,mota2024neutron,fabris_rastall,silva2023,mota2019combined}. 

In Ref.~\cite{Wojnar:2016bzk}, concerning the parameterization of the TOV (PTOV), the authors showed that it is possible to study various classes of extended theories of gravity by considering an effective energy-momentum tensor, which can be seen as the energy-momentum tensor of the current EoS corrected by the effects of a geometric source. What we denote here as generalized TOV equations (``GTOV'') was proposed in~\cite{clesio}, and this generalization includes an additional free {\it{ad hoc}} parameter in the PTOV mass term: $\Gamma\sqrt{\rho}M$, where $\Gamma$ is the new parameter, $\rho$ is the energy density and $M$ is the mass. The addition of this new parameter attempts to solve the hyperon puzzle~\cite{Chatterjee:2015pua} in NSs.
In Ref.~\cite{gtovprd}, the authors mapped the GTOV to the TOV in such a way as to include an anisotropic term and a modified energy density to the energy-momentum tensor. 
This mapping does not alter the results of the GTOV but it allows to calculate the mass-radius relation more straightforwardly. It also makes it easier to obtain the new equations to calculate the dimensionless tidal deformability with dependence on GTOV parameters.

Here we use Bayesian inference to find optimized values for the GTOV parameters in order to satisfy recent astrophysical constraints from GW detections and soft X-ray timing and spectroscopy, and using EoS parameterizations that have shown promising results in previous works. The big advantage of the approach adopted here, contrary to previous works on PTOV and GTOV~\cite{Wojnar:2016bzk,clesio,gtovprd} in which the authors had to limit themselves to vary only one or two parameters while keeping the others fixed, is that we can randomly vary all four parameters at the same time thousands of times and get a more complete view of their combined effect. Besides, in this work, we tried to optimize the parameters in such a way that we could satisfy at the same time both the constraints coming from the mass-radius relations and from the dimensionless tidal deformability of the NSs considered. 

We have applied uniform priors and, inspired by results obtained in~\cite{gtovprd}, set the intervals. For the likelihood distribution, we used a Gaussian distribution with constraints taken from NICER and GW events - more details in Sec.~\ref{sec_GTOV}. Additionally, we used four parameterizations of the eNLW, namely: Nl3$\omega\rho$, Nl3$\omega\rho$Y, El3$\omega\rho$, and El3$\omega\rho$Y, where Y stands for parameterizations with hyperons.

The paper is organized as follows. In Sec.~\ref{sec_eos} we present the EoSs used in this work. In Sec.~\ref{sec_GTOV} we briefly show the GTOV equations, the tidal deformability, and conditions for anisotropic stars. In Sec.~\ref{sec_res} we present the results and analysis. And in Sec.~\ref{sec_plus}, we execute some additional analysis of our results. Finally, in Sec.~\ref{conclusions} the conclusions are shown. 

\section{EoS Model for Neutron Star Matter} \label{sec_eos}

The nuclear equation of state (EoS) adopted in this work 
was based on the extended quantum hydrodynamics (QHD) model, which simulates the interaction between baryons 
through $\sigma\omega\rho$ mesons exchange. When hyperons are introduced, 
we include $\phi$ mesons with a hidden strangeness. The Lagrangian density is given by~\cite{2000csnp.book}:

\begin{align}
\mathcal{L}_{QHD} &= \sum_B \bar{\psi}_B[\gamma^\mu(\mbox{i}\partial_\mu  - g_{B\omega}\omega_\mu   - g_{B\rho} \frac{1}{2}\vec{\tau}_B \cdot \vec{\rho}_\mu) - (M_B - g_{B\sigma}\sigma)]\psi_B  -U(\sigma)   \nonumber   \\ 
&+ \frac{1}{2}(\partial_\mu \sigma \partial^\mu \sigma - m_\sigma^2\sigma^2) - \frac{1}{4}\Omega^{\mu \nu}\Omega_{\mu \nu} + \frac{1}{2} m_\omega^2 \omega_\mu \omega^\mu+ \frac{1}{2} m_\rho^2 \vec{\rho}_\mu \cdot \vec{\rho}^{ \; \mu} - \frac{1}{4}\vec{\bf{P}}^{\mu \nu} \cdot \vec{\bf{P}}_{\mu \nu} \label{s1}.
\end{align}
Here, the summation index $B$ runs over all the baryon octets present in the system, $\psi_B$ represents the baryonic fields with their masses $M_B$ and $\vec{\tau}$ are the Pauli matrices. The $\sigma,\, \omega^\mu$ and $\rho^\mu$ are meson fields, while $m_\sigma,\, m_\omega$ and $m_\rho$ are their corresponding masses. 
The $g_{Bi}$ (with $i = \sigma,\, \omega,\, {\rm or}  \ \rho   $) are the baryon-meson couplings that simulate the strong interaction.
The self-interaction potential 
is expressed as 
\begin{equation}
U(\sigma) =  \frac{\kappa M_N(g_{\sigma} \sigma)^3}{3} + \frac{\lambda(g_{\sigma}\sigma)^4}{4} \label{s2} ,
\end{equation}
 where $\kappa$, ${\lambda}$ and $g_{\sigma}$ are dimensionless coupling constants, $M_N$ is the nucleon mass and $\sigma$ has the dimensions of mass \cite{Boguta:1977xi}, chosen to be $1/fm$ which can be further converted into MeV through the $\hbar c= 197.33$ ~ MeV$\cdot$fm.
Aside from the non-linear $\sigma\omega\rho$ mesons that mediate the
baryonic
degrees of freedom, we introduce an additional vector meson $\phi$ with a hidden strangeness that mediates the interactions among the hyperons without influencing the properties of symmetric nuclear matter. 
Consequently, we introduce an additional Lagrangian density
\begin{equation}\label{El3} 
\mathcal{L}_\phi = -g_{Y\phi}\bar{\psi}_Y(\gamma^\mu\phi_\mu)\psi_Y + \frac{1}{2}m_\phi^2\phi_\mu\phi^\mu - \frac{1}{4}\Phi^{\mu\nu}\Phi_{\mu\nu} , 
\end{equation}
as presented in~\cite{Lopes:2020rqn, Lopes:2019shs, Weissenborn:2011kb}, which is important to obtain massive neutron stars. In addition, we introduce the $\omega$-$\rho$ coupling in a similar way as presented in 
~\cite{Fattoyev:2010mx}:
\begin{equation}
 \mathcal{L}_{\omega\rho} = \Lambda_{\omega\rho}(g_{\rho}^2 \vec{\rho^\mu} \cdot \vec{\rho_\mu}) (g_{\omega}^2 \omega^\mu \omega_\mu) ,
\end{equation}
which is necessary to fix the symmetry energy slope ($L$) that significantly affects the radii and the tidal deformability of the stars \cite{Cavagnoli:2011ft, Dexheimer:2018dhb}. The couplings and parameters of the Lagrangians presented above are in subsection A, table I and table II.

We introduce leptons as free Fermi gas 
to obtain $\beta$-equilibrated charge-neutral stellar matter. The detailed calculation of $\beta$-equilibrium EoS for symmetric nuclear matter has been extensively studied in QHD formalism in the literature (see \textit{e.g.} \cite{2000csnp.book, 1992RPPh...55.1855S, Menezes:2021jmw}) and we do not intend to repeat them here. In the same vein, the calculation of the six nuclear parameters at the saturation density ($n_0,~ M_N^*/M_N,~K,~S_0,~L,~B/A$) can be 
verified in \cite{2000csnp.book, Cavagnoli:2011ft} and the references therein. {The pressure (P) and the energy density ($\varepsilon$) contributions that constitute the EoS of the system can be determined from the energy-momentum tensor ($T_{\mu\nu}$), yielding 
\begin{equation}
    \varepsilon =\langle T_{00}\rangle \quad{\rm and}\quad P = \dfrac{1}{3}\langle T_{jj}\rangle,
\end{equation}
where $\langle T_{00}\rangle $ is the time component of the $T_{\mu\nu}$ trace, representing the energy density and $\langle T_{jj}\rangle$ is the spatial component of the $T_{\mu\nu}$ trace, representing pressure. The detailed calculations of the EoS have been widely documented in the literature and we do not intend to repeat them here, interested readers can find a review in \cite{Menezes:2021jmw} (and references therein). 
}

\subsection{Coupling constraints}

When hyperons are included in the stellar matter, it is important to determine the strength of the hyperon-meson coupling constants. Except for the $\Lambda^0$ coupling which is well known with a potential depth $U_\Lambda=-28~ {\rm MeV}$, the coupling strength of the rest of the hyperons is calculated using different approaches in the literature. Some of the methods adopted are: the universal coupling \cite{PhysRevLett.67.2414}, fixed \cite{PhysRevC.85.065802} and nonfixed \cite{PhysRevC.89.025805, Weissenborn:2011kb, PhysRevC.88.015802} potential depths, symmetry argument and non-symmetry arguments  \cite{Lopes:2022vjx, PhysRevC.84.065810}. We follow the formalism present in \cite{Lopes:2020rqn,Lopes:2022vjx,lopes2022nature} to study hyperonic neutron stars assuming that the vector mesons are constrained through the SU(3) symmetry group and the scalar mesons are constrained by fixing the potentials $U_\Sigma$ and $U_\Xi$, giving rise to the hyperon-meson constraints:
\begin{equation}
    \dfrac{g_{\Lambda\omega}}{g_{N\omega}} = \dfrac{4+2\alpha_v}{5+4\alpha_v},\quad\quad  \dfrac{g_{\Sigma\omega}}{g_{N\omega}} = \dfrac{8-2\alpha_v}{5+4\alpha_v}, \quad{\rm and}\quad \dfrac{g_{\Xi\omega}}{g_{N\omega}} = \dfrac{5-2\alpha_v}{5+4\alpha_v},
\end{equation}
for hyperon and the $\omega$ meson couplings
\begin{equation}
    \dfrac{g_{\Lambda\phi}}{g_{N\omega}} = \sqrt{2}\Bigg(\dfrac{2\alpha_v -5}{5+4\alpha_v}\Bigg), \quad\quad \dfrac{g_{\Sigma\phi}}{g_{N\omega}} = \sqrt{2}\Bigg(\dfrac{-2\alpha_v -1 }{5+4\alpha_v}\Bigg), \quad{\rm and}\quad \dfrac{g_{\Xi\phi}}{g_{N\omega}} = \sqrt{2}\Bigg(\dfrac{-2\alpha_v -4 }{5+4\alpha_v}\Bigg),
\end{equation}
for hyperon and $\phi$ meson couplings
\begin{equation}
    \dfrac{g_{\Lambda\rho}}{g_{N\rho}} = 0, \quad\quad \dfrac{g_{\Sigma\rho}}{g_{N\rho}} =2\alpha_v, \quad{\rm and}\quad \dfrac{g_{\Xi\rho}}{g_{N\rho}} =-( 1-2\alpha_v).
\end{equation}
for hyperon $\rho$ meson couplings. 
We must note that $g_{N\phi}$= 0.
In the current work we assume $\alpha_v = 1$, 
which recovers the SU(6) parameterization for vector mesons \cite{RevModPhys.38.215}. Besides, we assume potentials $U_\Sigma =30\,\rm MeV$ and $U_\Xi = - 4 \,\rm MeV$ for the El3$\omega\rho$ and Nl3$\omega\rho$ models, with $U$ being the potential depth and its subscripts representing the particle under consideration.
We can infer from the works in \cite{Lopes:2020rqn,Lopes:2022vjx,lopes2022nature} that the smaller the value of $\alpha_v$, the higher the maximum stellar mass. We used two sets of parameterizations for the QHD model, with and without hyperons, the so-called Nl3* (we refer to it as Nl3$\omega\rho$) and the modified Nl3* (we refer to it as the El3$\omega\rho$) as presented in Tab.~\ref{T3}. 
The masses of the baryon octet are: $M_N = 939$ MeV, $M_\Lambda = 1116$ MeV, $M_\Sigma = 1193$ MeV, and $M_\Xi = 1318$ MeV and the leptons masses are: $m_e = 0.51$ MeV and $m_\mu = 105.6$ MeV, where subscripts represent the type of particle under consideration.

\begin{table}[!ht]
\begin{center}
\begin{tabular}{|c |c| c| c|c|}
\hline
  Parameters & Values & Parameters & Values & Meson masses [MeV]\\
 \hline
  \multicolumn{5}{|c|}{Nl3$\omega\rho$ parameterization}\\
  \hline
$g_{N\sigma}$& 10.094 & $n_0\,{\rm [fm}^{-3}]$& 0.150& $m_\sigma = 508.194$\\  
$g_{N\omega}$ & 12.807 & $M^*/M$ &0.594  & $m_\omega = 782.501$\\
 $g_{N\rho}$& 14.441 & $K\,[{\rm MeV}]$& 258 & $m_\rho = 763$\\
 $\lambda$& -0.002904 & $S_0[{\rm MeV}]$& 30.7& $m_\phi = 1020$\\
 $\kappa$ &-0.002208 & $L\,[{\rm MeV}]$& 42& --\\
  $\Lambda_{\omega\rho}$ & 0.045 & $B/A\,[{\rm MeV}]$ & 16.31&--\\
  
 \hline
  \multicolumn{5}{|c|}{El3$\omega\rho$ parameterization}\\
  \hline
$(g_{N\sigma}/m_\sigma)^2$& 12.108$\,[{\rm fm}^2]$ & $n_0\,{\rm [fm}^{-3}]$& 0.156& $m_\sigma = 512$ \\  
$(g_{N\omega}/m_\omega)^2$ & 7.132$\,[{\rm fm}^2]$ & $M^*/M$ & 0.69 & $m_\omega = 783$\\
 $(g_{N\rho}/m_\rho)^2$&5.85$\,[{\rm fm}^2]$ & $K\,[{\rm MeV}]$& 256& $m_\rho = 770$\\
 $\kappa$&0.004138& $S_0[{\rm MeV}]$& 32.1 &  $m_\phi = 1020$\\
 $\lambda$ &{-0.00390} & $L\,[{\rm MeV}]$& 66  &--\\
  $\Lambda_{\omega\rho}$ & 0.0185 & $B/A\,[{\rm MeV}]$ & 16.2&--\\
 \hline
\end{tabular}
\caption{Nl3$\omega\rho$ and El3$\omega\rho$ parameterizations \cite{biesdorf2023qcd}. The phenomenological constraints for the symmetric nuclear matter at saturation density were derived from \cite{Dutra:2014qga, Oertel:2016bki, Li:2021thg, LIGOScientific:2018cki} and \cite{Reed:2021nqk}.} 
\label{T3}
\end{center}
\end{table}

\begin{table}[h!]
\centering

\begin{tabular}{|c|c|c|c|}
\hline
Model & $g_{\Lambda\sigma}/g_{N\sigma}$ & $g_{\Sigma\sigma}/g_{N\sigma}$ & $g_{\Xi\sigma}/g_{N\sigma}$ \\ \hline
NL3$\omega\rho$ & 0.613 & 0.461 & 0.279 \\ 
EL3$\omega\rho$ & 0.610 & 0.406 & 0.269 \\ \hline
\end{tabular}
\caption{Hyperon-$\sigma$ coupling constants adjusted to reproduce the hyperon potential in strange neutron matter derived from hypernuclear observables \cite{biesdorf2023qcd}.}
\label{T4}
\end{table}

\section{Generalized Tolman-Oppenheimer-Volkoff Equation (GTOV)} \label{sec_GTOV}

In this section, we discuss the generalized hydrostatic equilibrium equations and show how to construct these modified expressions using the Einstein field equation with an effective anisotropic stress-energy tensor. Usual TOV equations extend the Newtonian hydrostatic equilibrium equation to include both special relativistic and general relativistic effects. In this sense, it can be inferred that modifications in general relativity would lead to extra terms in the TOV equations. Taking into account the compactness $C=G M/(R c^2)$ of a star, we are interested in NSs where $ 1/6 < C  < 1/3 $, \textit{i.e.} compact objects~\cite{Iyer_1985}. As a consequence, general relativistic corrections associated with the curvature of the spacetime or even effects of modified theories of gravity, are expected in this regime. Let us start by writing the GTOV in the following form with $(G=c=1)$ and $(+,-,-,-)$ signature
\begin{equation}
    \frac{dP}{dr}=-\frac{(1 + \alpha)(\varepsilon + \beta P)(M + 4 \chi \pi r^3 P)}{r (r - 2 M)},
    \label{3eq1}
\end{equation}
\begin{equation}
    \frac{dM}{dr}=4 \pi r^2 (\varepsilon + \theta P) + \Gamma \sqrt{\varepsilon} M.
    \label{3eq2} 
\end{equation}
It is worth noting that the parameters $\alpha,\beta,\chi,\theta$ and $\Gamma$ in equations (\ref{3eq1}) and (\ref{3eq2}) generalize standard results in GR and have particular physical meaning.
\begin{itemize}
\item The term $\alpha$ can be associated with an effective gravitational coupling $G_{eff}=(1+\alpha)$, where $G_{eff}=1$ is the standard gravitational coupling. In this way, $\alpha$ can be seen as an effect of modified theories of gravity. For example, the Einstein-Hilbert action in $f(R)$ gravity can be generalized to be a function of $R$, where $f(R)$ represents a function of the Ricci scalar $R$. Specifically, a commonly used function in $f(R)$ is written in the form $f(R)= R + \mu R^{n}$, with $\mu$ the constant parameter that controls the contribution of the higher-order term in $R$. In particular, for $f(R)$ gravity, $\alpha=1/3$~\cite{brax2008f}. 
\item The parameter $\beta$ is an effect of the inertial pressure since it modifies the usual term $(\varepsilon + P)$ originating from the hydrostatic equilibrium equation $\nabla_{\mu}T^{\nu\mu}=0$ in spherical coordinates to the form $(\varepsilon + \beta P)$. We can see that the absolute values of this parameter increase the value of the right side of the hydrostatic equilibrium equation, \textit{i. e.}, $|\beta| > 1$, and consequently the variation of pressure in relation to the radial position. The GR values are obtained by setting $\beta = 1$.
\item  $\chi$ parameterizes the self-gravity of pressure in compact stars. This type of influence is strictly non-Newtonian, which indicates that the value $\chi=1$ fully captures the effect of GR \cite{schwab2008self,Rappaport:2007ct}. Values corresponding to $\chi \rightarrow 0$ remove the effect of self-gravity of pressure from the equations. 
It is believed that in addition to the mass-radius relations of NSs, in the cosmological context the self-gravity of pressure could be detectable.
\item $\theta$ takes into account possible changes in the mass function due to the gravitational effects of pressure. Thus, this term modifies the function $M(r)$ by adding an extra quantity to the formula for the mass.
\item As for the parameter $\Gamma$, it was introduced in~\cite{clesio} to capture additional effects of the modified hydrostatic equilibrium equations. As we can see in Eq. (\ref{3eq2}), $\Gamma$ parameterizes a new coupling between energy density and the mass function. This parameter was counted as second order correction in the continuity equation for the mass (\ref{3eq2}). 
\end{itemize}
As expected, TOV in GR is recovered considering particular values for the parameters: $\alpha=0,\;\beta=1,\;\chi=1,\;\theta=0$ and $\Gamma=0$. In addition, some values for GTOV  parameters associated with modified gravity can be used in order to take into account the influence of such theories. 
The possibility of mapping modified theories of gravity into GR with an effective energy-momentum tensor is well-known in the literature. For example, some non-conservative extensions of GR \cite{rastall1972generalization,mota2019combined} carry extra terms in the field equations that can be manipulated to be included in the energy-momentum tensor, providing GR with an effective energy-momentum tensor. In general, the field equations associated with extended theories of gravity can be defined in the form \cite{Wojnar:2016bzk}
\begin{equation}
\gamma (\Psi^i)(G_{\mu\nu} - W_{\mu\nu})=kT_{\mu\nu},
\label{3eq15}
\end{equation}
where $k=8\pi$ and $G_{\mu\nu}$ the Einstein tensor. As we can see, the effects of modified gravity are represented by the extra terms $\gamma (\Psi^i)$ and $W_{\mu\nu}$. The term $W_{\mu\nu}$ represents additional geometric effects of modified gravity while $\gamma (\Psi^i)$ encodes the possibility of modifying the coupling with the matter fields where $\Psi^i$ represents either additional gravitational fields or curvature invariants. It is easy to see that equation (\ref{3eq15}) can be written in the following form
\begin{equation}
G_{\mu\nu} =kT_{\mu\nu}^{\text{eff}}=\frac{k}{\gamma}T_{\mu\nu}+W_{\mu\nu}.
\label{3eq16}
\end{equation}
It follows from the Bianchi identity that $\nabla^{\mu} T_{\mu\nu}^{\text{eff}}=0$, \textit{i.e.}, $T_{\mu\nu}^{\text{eff}}$ is conserved. Notice that it is possible to derive the hydrostatic equilibrium equation for a large class of modified gravities obtained from the field equation (\ref{3eq16}). As discussed in \cite{Wojnar:2016bzk}, the generalized TOV in this case has a more complicated form than the GTOV equation. For instance, in the case of Rastall gravity \cite{rastall1972generalization} the Bianchi identity gives the expression $\alpha_1 \varepsilon + \alpha_2 P$, where $\alpha_1$ and $\alpha_2$ are constants. By comparing these relations with Eq. (\ref{3eq1}), we conclude that the term $\beta$ in GTOV parameterizes the influence of constant $\alpha_2$. In this way, some effects of the TOV associated with modified theories of gravities can be captured by the parameters used in GTOV.  Then the effective influence on the stellar structure can be studied by appropriately varying the GTOV parameters. In the following section, we analyze how GTOV can be connected with usual TOV in GR using an effective anisotropic energy-momentum tensor.

\subsection{Connecting anisotropy in GR}\label{connections}

Initially, GTOV was proposed as an \textit{ad hoc} parameterization of the TOV~\cite{clesio}. It was recently discovered that the GTOV can be constructed considering an anisotropic stress-energy tensor in GR~\cite{gtovprd}. In this case, we consider the general expression for the anisotropic fluid defined by
\cite{Arbanil2016,Garattini2016,Sharif2018}:
\begin{equation}
    T_{\mu\nu}=Fg_{\mu\nu}+(F+\Bar\rho)U_{\mu}U_{\nu}+(P-F)N_{\mu}N_{\nu}
    \label{3eq3},
\end{equation}
where $F$, $\Bar\rho$ and $P$ are the transverse pressure, the energy density and the radial pressure of the fluid respectively. The vectors $U_{\mu}$ and $N_{\mu}$ represent respectively the four velocity and radial unit vector, defined by
 \begin{equation}
        U^{\mu}=\left( \frac{1}{\sqrt{B}},0,0,0\right)
    \label{3eq4},
 \end{equation}
 \begin{equation}
        N^{\mu}=\left( 0,\frac{1}{\sqrt{A}},0,0\right)
    \label{3eq5},
 \end{equation}
and that obey the conditions: $U_{\nu}U^{\nu}=-1$, $N_{\nu}N^{\nu}=1$ and $U_{\nu}N^{\nu}=0$.
Considering a spherically symmetric spacetime 
\begin{equation}
    ds^{2}=B dt^{2}-A dr^{2}-r^{2}(d\theta^{2}+\sin{\theta}^{2}d\phi^{2}),
    \label{3eq6}
\end{equation}
where $A$ and $B$ are undetermined radial functions and the anisotropic energy-momentum tensor, the Einstein field equations can be written as  
\begin{equation}
 -\frac{B}{r^{2}A} + \frac{B}{r^{2}} + \frac{A'B}{rA^{2}} = 8\pi B \Bar{\rho},  \label{3eq7}
\end{equation}
\begin{equation}
-\frac{A}{r^{2}} + \frac{B'}{rB} + \frac{1}{r^2}  = 8\pi A P,  \label{3eq8}
\end{equation}
\begin{align}
 -\frac{B'^{2}r^{2}}{4AB^{2}} - \frac{A'B'r^{2}}{4A^{2}B} + \frac{B'' r^{2}}{2AB} - \frac{A'r}{2A^{2}} + \frac{B'r}{2AB}  
 = 8\pi r^{2} F,  \label{3eq9}
\end{align}
where the prime ($'$) denotes the differentiation with respect to the radial coordinate. Standard manipulations of the equations (\ref{3eq7}), (\ref{3eq8}) and (\ref{3eq9}) taking into account the conservation equation for the fluid, give the result
\begin{equation}
P' = -(P+\Bar{\rho})\frac{M+4\pi r^{3}P}{r(r-2M)} - \frac{2}{r}\sigma, 
\label{3eq10}
\end{equation}
and 
\begin{equation}
    M'=4\pi \Bar{\rho} r^2.
    \label{3eq11}
\end{equation}
These expressions describe the hydrostatic equilibrium in the context of general relativity. The presence of anisotropy introduces the extra term $2\sigma/r$, where $\sigma$ is the anisotropy factor, and is defined as  
 \begin{equation}
    \sigma \equiv P- F,
    \label{3eq12}
\end{equation}
where $P$ is the radial pressure and $F$ is the transverse pressure. By defining the particular form of $\Bar{\rho}$ and $\sigma$ in equations (\ref{3eq10}) and (\ref{3eq11}) as in~\cite{gtovprd}
\begin{equation}
    \Bar{\rho}=\rho + \theta P + \frac{\Gamma M \sqrt{\rho}}{4\pi r^2},
    \label{3eq13}
\end{equation}
and
\begin{align}
    \sigma=&\frac{ \left(M+4 \pi r^{3} P\right) (P(1+\theta)+\rho ) \left[(\alpha +1) \left(\frac{(\chi -1) 4 \pi r^{3} P}{M+4 \pi r^{3} P}+1\right) \left(\frac{P (\beta -\theta -1)}{P(1+\theta)+\rho }+1\right)-1\right]}{2 (r-2  M)}\\
    & - \frac{ \Gamma  M \sqrt{\rho } \left(M+4 \pi r^{3} P\right)}{8 \pi  r^2(r-2  M)},
    \label{3eq14}
\end{align}   
we obtain the GTOV given by Equations (\ref{3eq1}) and (\ref{3eq2}). It should be noted that 
 one specific choice for what is known as
the generalized TOV (GTOV)~\cite{Gokhroo:1994fbj, Dev:2000gt} in the literature is given by  $P' = -(P+\rho)\frac{M+4\pi r^{3}P}{r(r-2M)} - \frac{2}{r}\sigma, 
$  and     $M'=4\pi \rho r^2$  
where negative sign in front of $\sigma$ depends on the definition of this term. Some articles use $\sigma \equiv F-P$. In the next sections, we discuss the effects of GTOV on the physical structure of NSs. It is important to note that we also used $\Gamma=0$ in the same way as the Ref.~\cite{gtovprd}, as we used the same GTOV parameter intervals in the priors.

For an anisotropic NS with a perfect fluid, the following conditions must be satisfied~\cite{Estevez-Delgado:2018ydn,Setiawan:2019ojj}:
\begin{enumerate}\label{conditions_ani}
    \item Inside the star we must have $F>0$, $P>0$ and $\rho >0 $;
    \item The gradient of $\rho$ and $P$ must be monotonically decreasing, $\frac{d\rho}{dr} < 0$ , $\frac{dP}{dr} <0$ and the maximum value must be at the center;
    \item Three conditions for the anisotropic fluid must be satisfied within the star: a) null energy ($\rho >0$),  b) dominant energy ($\rho+ P >0$, $\rho+F>0$), and c) strong energy ($\rho+P + 2F > 0$);
    \item Both transverse and radial  speeds of sound must be causal inside the star, \textit{i.e.},   0 < $\frac{\partial P}{\partial \rho} <1 $ and $0<\frac{\partial F}{\partial \rho}$  < 1, where $c_s^2(radial)=\frac{\partial P}{\partial \rho}$  and $c_s^2(transverse)=\frac{\partial F}{\partial \rho}$;
    \item Both transverse and radial pressure must be the same at the center of the star, $P(0)=F(0)$. 
\end{enumerate}

\subsection{Tidal deformability for anisotropic NS}

For a binary system, the shape of each star is deformed due to the external field ($\epsilon_{ij}$) of its companion. Therefore, the stars develop a quadrupole moment ($Q_{ij}$). The quadrupole moment has a linear dependence on the tidal field ($\lambda$), which is given by,
\begin{equation}
    Q_{ij}=-\lambda\epsilon_{ij}.
\end{equation}
The dimensionless tidal deformability ($\Lambda$) is related with the dimensionless compactness parameter $C \equiv M/R$ as:
\begin{equation}
  \Lambda \equiv \frac{\lambda}{M^{5}} \equiv \frac{2 k_{2}}{3 C^{5}}.  
\end{equation}
The tidal deformability parameter quantifies how easily the compact object -- in our case, the NS-NS binary merger -- is deformed when subject to an external field ($\epsilon_{ij}$). The higher the $\lambda$ value, the more deformable the compact object. The second order (quadrupole) Love number $k_{2}$~\cite{Hinderer:2007mb} is found using the following expression: 
\begin{equation}
\begin{aligned}
k_2= & \frac{8 C^5}{5}(1-2 C)^2\left[2-y+2 C\left(y-1\right)\right] \\
& \times\left\{2 C\left[6-3 y+3 C\left(5 y-8\right)\right]\right. \\
& +4 C^3\left[13-11 y+C\left(3 y-2\right)+2 C^2\left(1+y\right)\right] \\
& \left.+3(1-2 C)^2\left[2-y+2 C\left(y-1\right)\right] \log (1-2 C)\right\}^{-1}.
\end{aligned}
\end{equation}
To obtain the value of $y$, we must solve the following first-order differential equation together with the GTOV:
\begin{equation}\label{y}
    r\frac{d y}{d r} + y^{2} + y B_1 + r^{2} B_2=0,
\end{equation}
where
\begin{equation}
     B_1=\frac{r - 4 \pi r^{3} (\Bar{\rho}-P)}{r-2 M},
\end{equation}
and
\begin{equation}
    B_2=\frac{4 \pi r\left[4\Bar{\rho}+8 P+\frac{(\Bar{\rho}+P)(1+d\Bar{\rho}/dP)}{1-d\sigma/dP}-\frac{6}{4 \pi r^{2}}+4\sigma\right]}{r-2 M}-4 \left[\frac {M+4 \pi r^{3} P}{r^{2}(1- 2 M/r)}\right]^{2}.
\end{equation}
The boundary condition to solve Eq. (\ref{y}) at $r=0$ is given by $y(0)=2$. The isotropic case~\cite{Postnikov:2010yn} is recovered by setting  $\sigma=0$, and $\bar{\rho} = \rho$. Further, the term $d\sigma/dP$ was obtained analytically. For the numerical derivation of $c_s^2(radial)$, we used central differences for the central values of the tabulated EoS and finite differences at the boundaries.

\section{Results and Analysis} \label{sec_res}

%%%%%%%%%%%%%%%%%%%%%%%%%%%%%%%%%%%%%%%%%%%%%%%%%%%%%%%%%%%%%%%%%%%%%%%%%%%%%%%%%%%%
\begin{figure}[!ht] 
    \centering
        \includegraphics[width=0.51\textwidth]{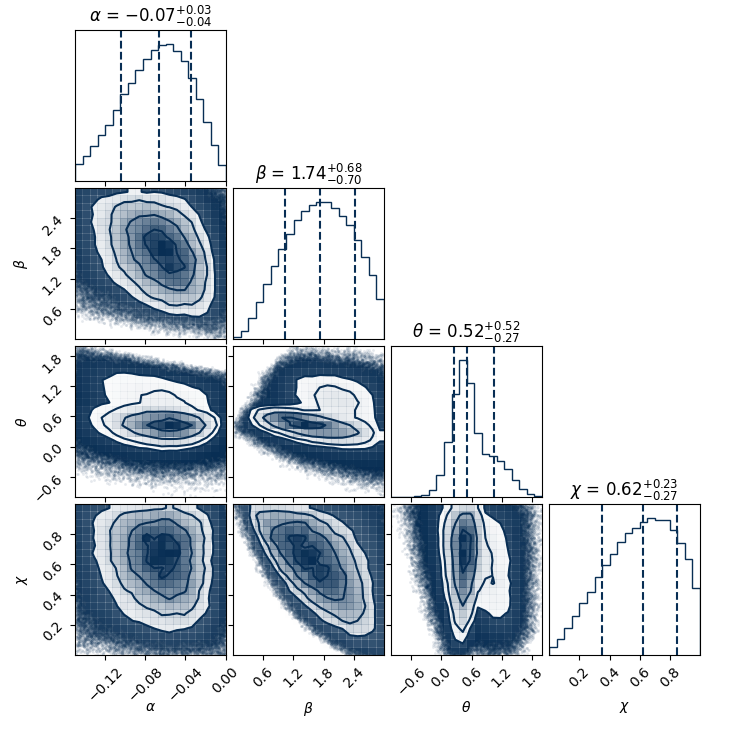}
        \includegraphics[width=0.49\textwidth]{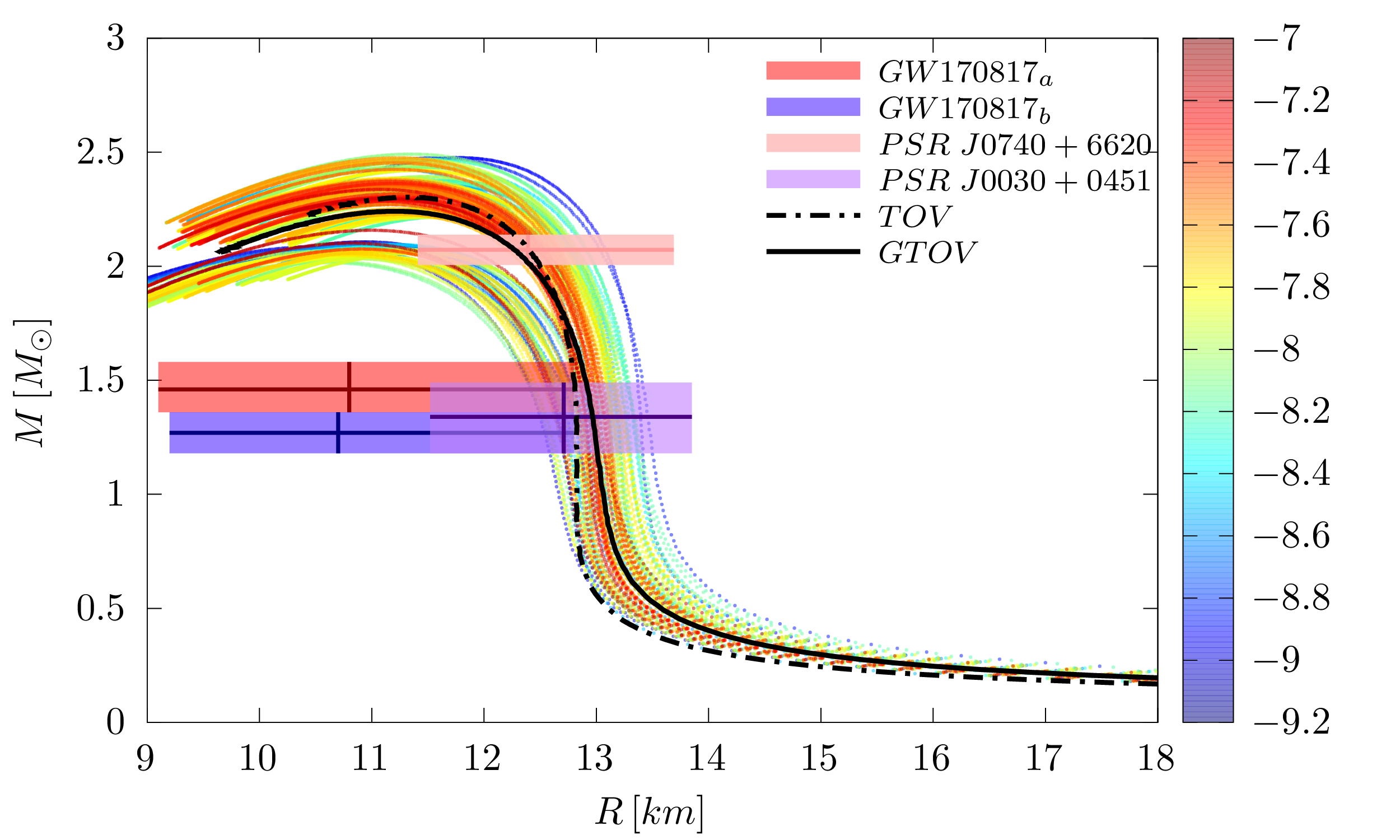}
        \includegraphics[width=0.49\textwidth]{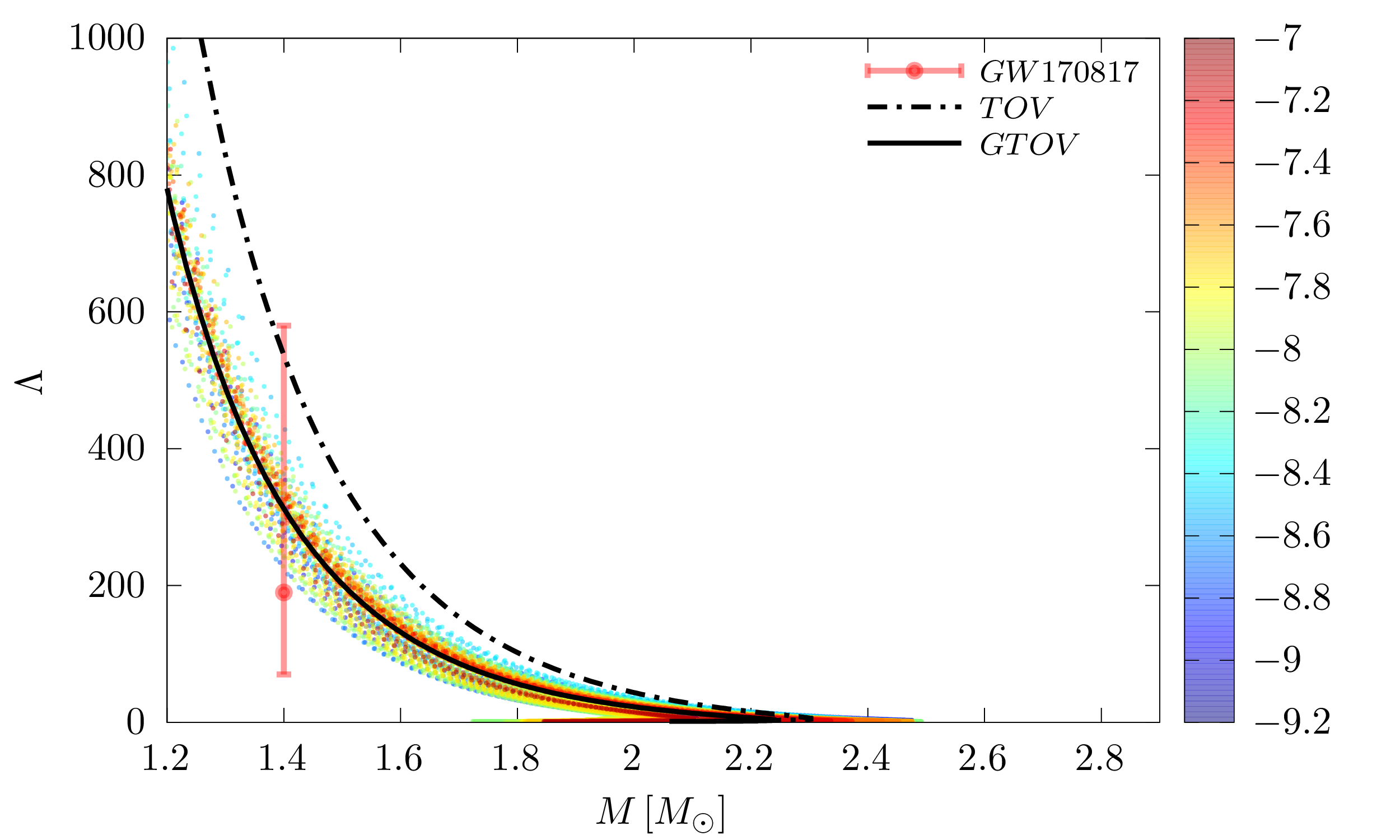}
    \caption{On top we show the corner plot of the posterior distribution of the GTOV parameters, on bottom left we show the mass-radius diagram and on bottom right we show the tidal deformability as a function of the mass. All plots are for El3$\omega\rho$ EoS in the conservative scenario.} \label{el3wp_cons}
\end{figure}
%%%%%%%%%%%%%%%%%%%%%%%%%%%%%%%%%%%%%%%%%%%%%%%%%%%%%%%%%%%%%%%%%%%%%%%%%%%%%%%%%%%%

%%%%%%%%%%%%%%%%%%%%%%%%%%%%%%%%%%%%%%%%%%%%%%%%%%%%%%%%%%%%%%%%%%%%%%%%%%%%%%%%%%%%
\begin{figure}[!ht] 
    \centering
        \includegraphics[width=0.51\textwidth]{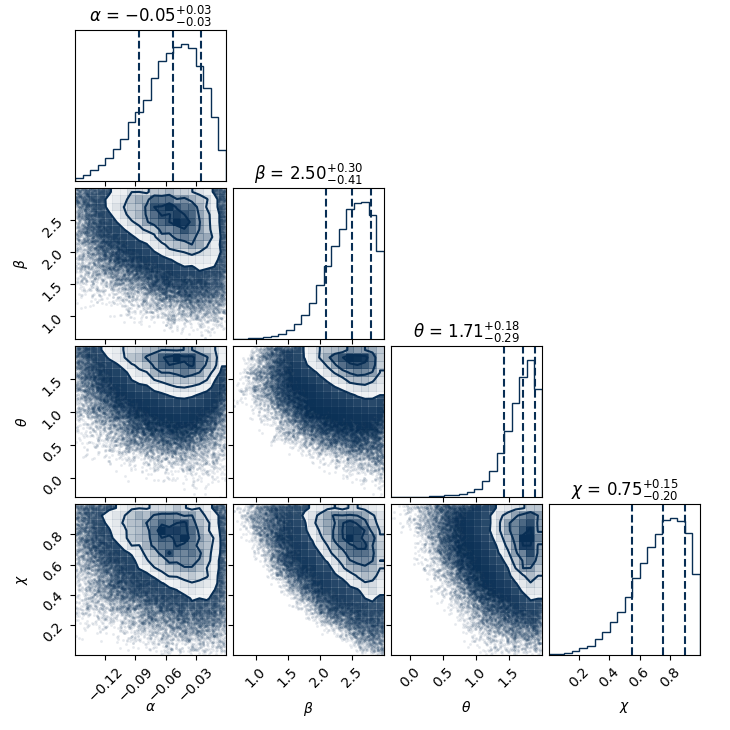}
        \includegraphics[width=0.49\textwidth]{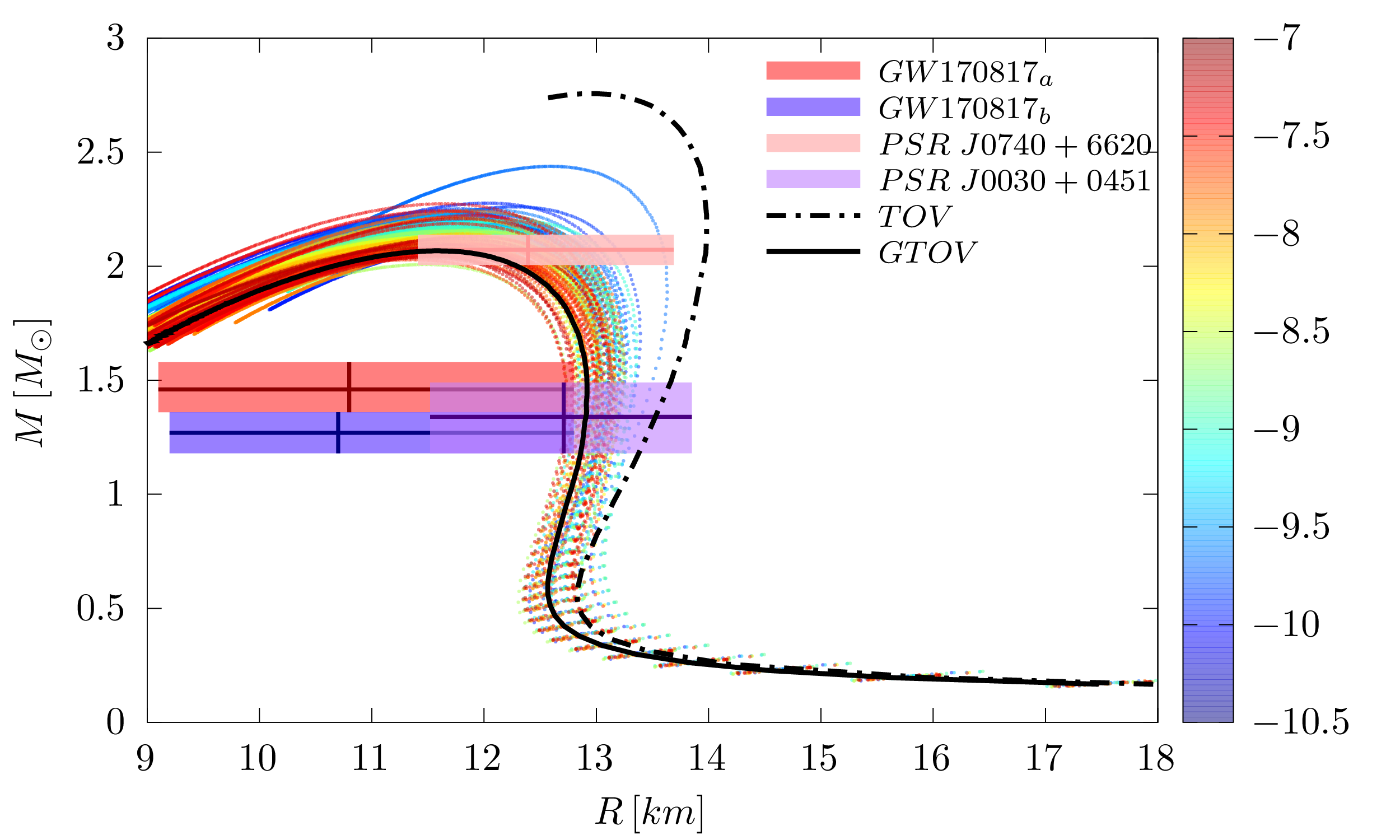}
        \includegraphics[width=0.49\textwidth]{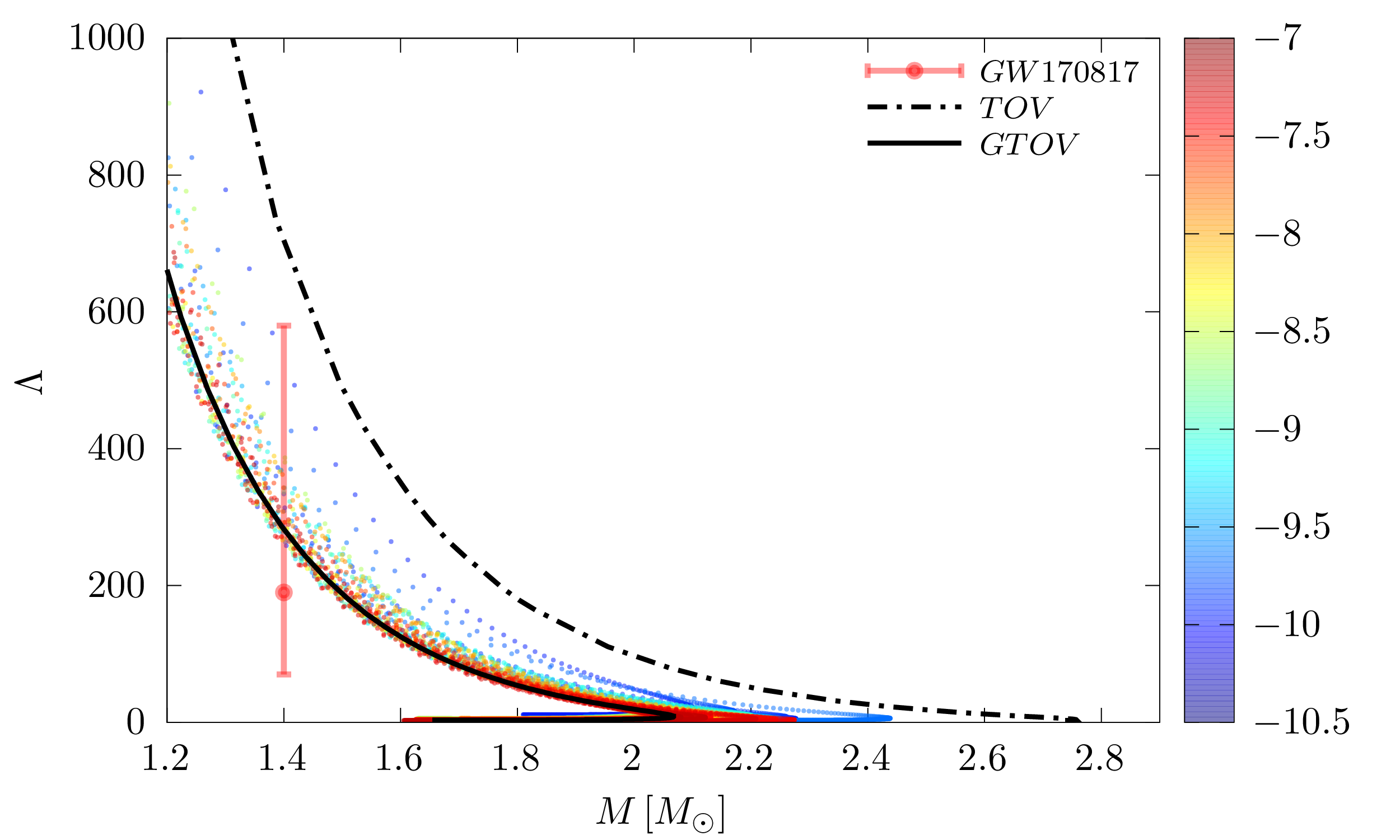}
    \caption{On top we show the corner plot of the posterior distribution of the GTOV parameters, on bottom left we show the mass-radius diagram and on bottom right we show the tidal deformability as a function of the mass. All plots are for Nl3$\omega\rho$ EoS in the conservative scenario.} \label{nl3wp_cons}
\end{figure} 
%%%%%%%%%%%%%%%%%%%%%%%%%%%%%%%%%%%%%%%%%%%%%%%%%%%%%%%%%%%%%%%%%%%%%%%%%%%%%%%%%%%%

%%%%%%%%%%%%%%%%%%%%%%%%%%%%%%%%%%%%%%%%%%%%%%%%%%%%%%%%%%%%%%%%%%%%%%%%%%%%%%%%%%%%
\begin{figure}[!ht]
    \centering
        \includegraphics[width=0.51\textwidth]{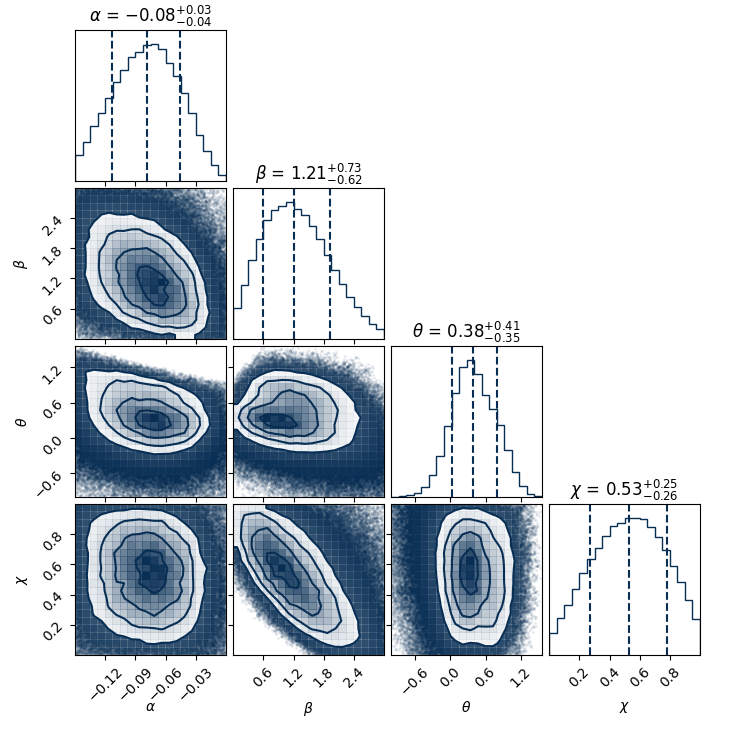}
        \includegraphics[width=0.49\textwidth]{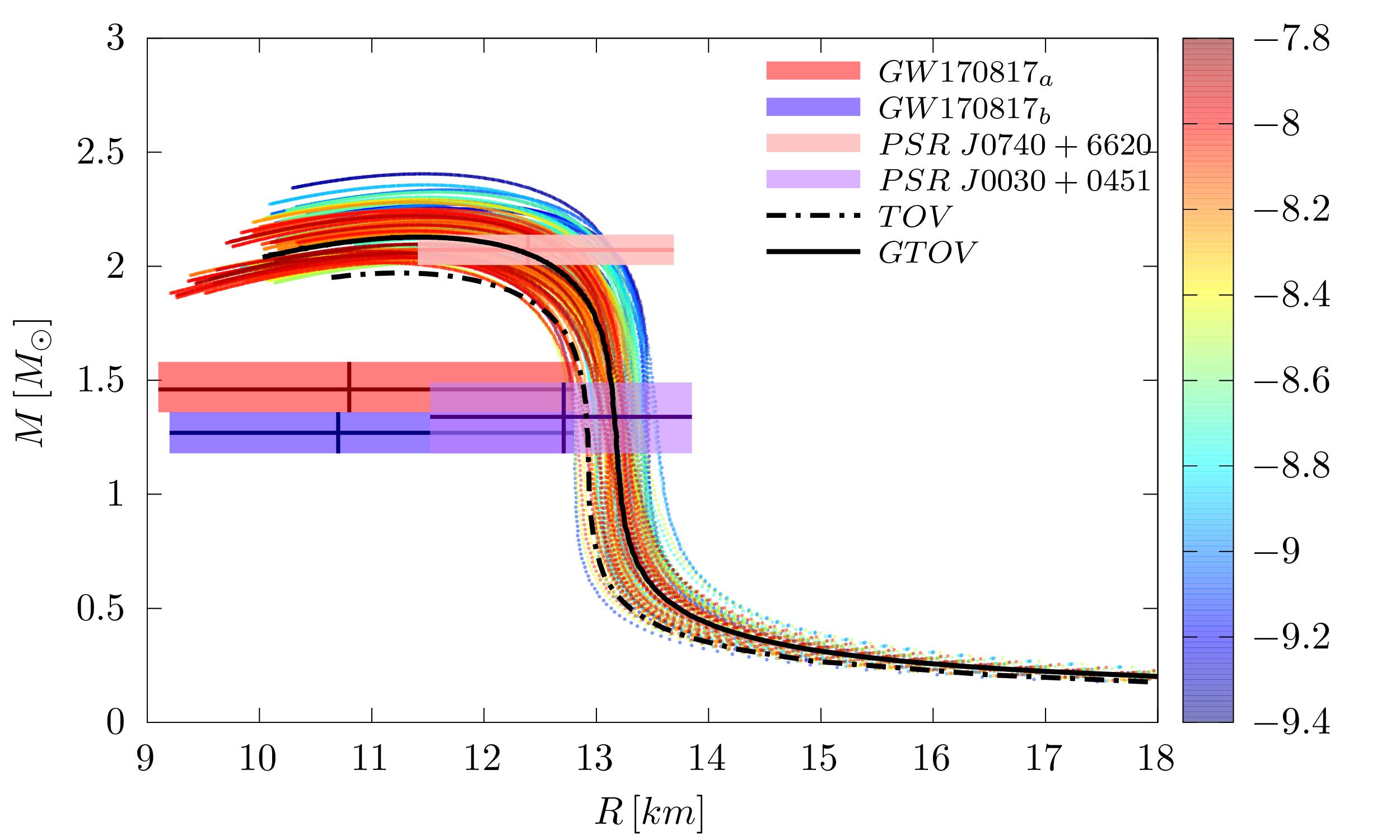}
        \includegraphics[width=0.49\textwidth]{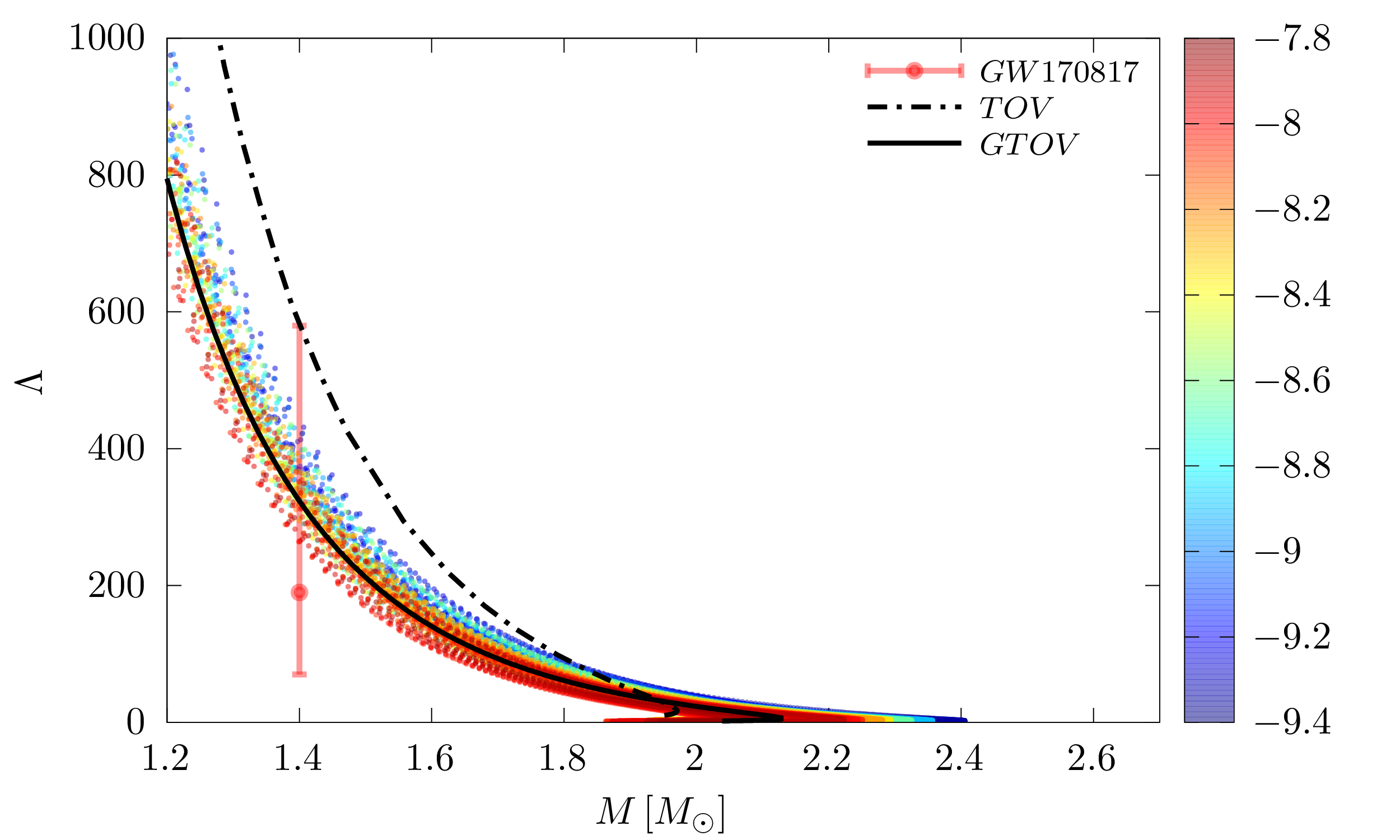}
    \caption{On top we show the corner plot of the posterior distribution of the GTOV parameters, on bottom left we show the mass-radius diagram and on bottom right we show the tidal deformability as a function of the mass. All plots are for El3$\omega\rho$Y EoS in the conservative scenario.} \label{el3wpy_cons}
\end{figure}
%%%%%%%%%%%%%%%%%%%%%%%%%%%%%%%%%%%%%%%%%%%%%%%%%%%%%%%%%%%%%%%%%%%%%%%%%%%%%%%%%%%%

%%%%%%%%%%%%%%%%%%%%%%%%%%%%%%%%%%%%%%%%%%%%%%%%%%%%%%%%%%%%%%%%%%%%%%%%%%%%%%%%%%%%
\begin{figure}[!ht]
    \centering
        \includegraphics[width=0.51\textwidth]{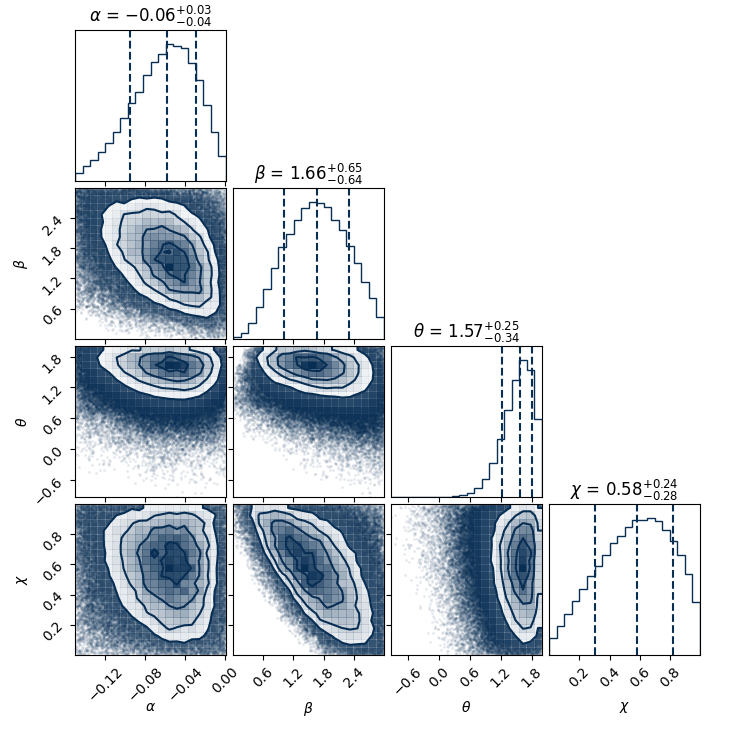}
        \includegraphics[width=0.49\textwidth]{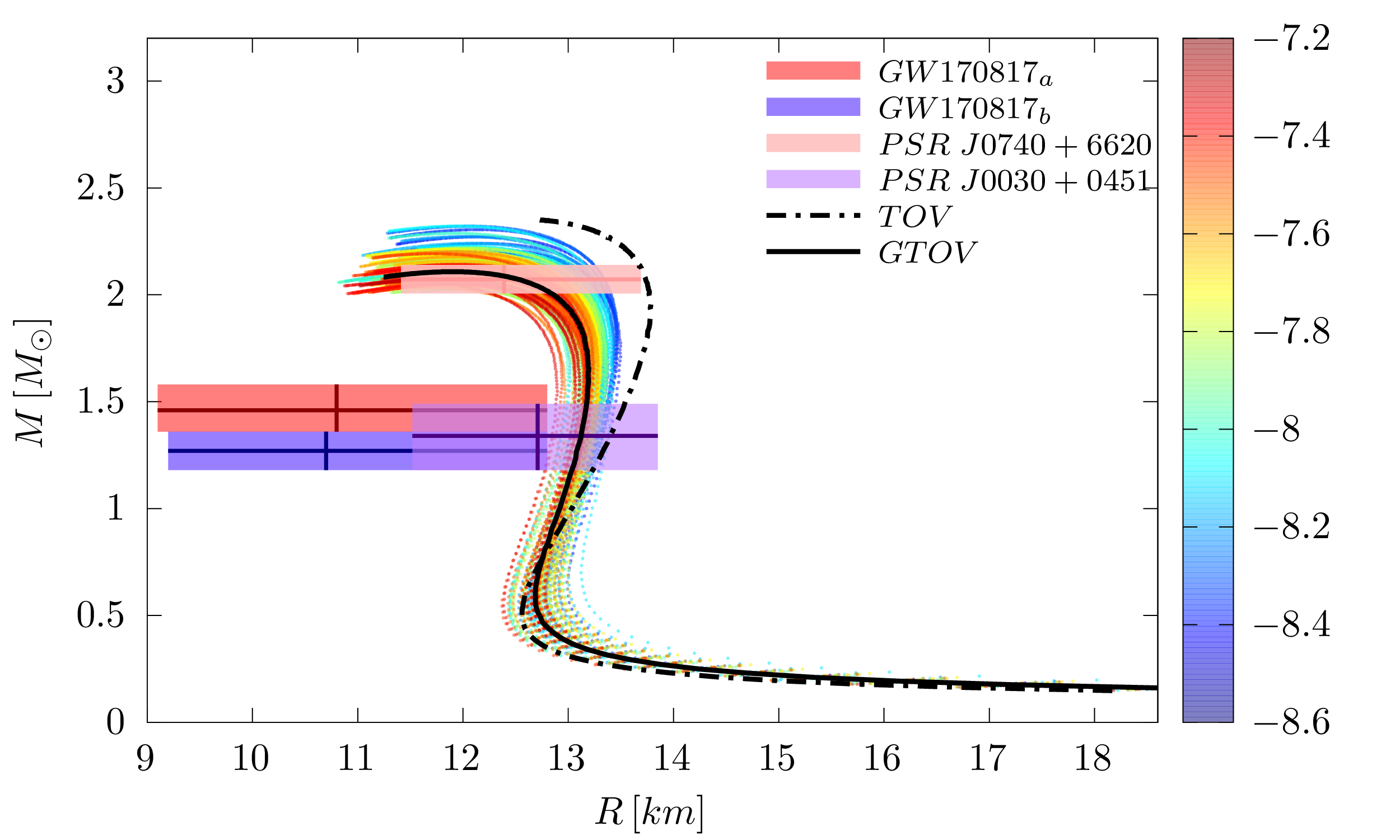}
        \includegraphics[width=0.49\textwidth]{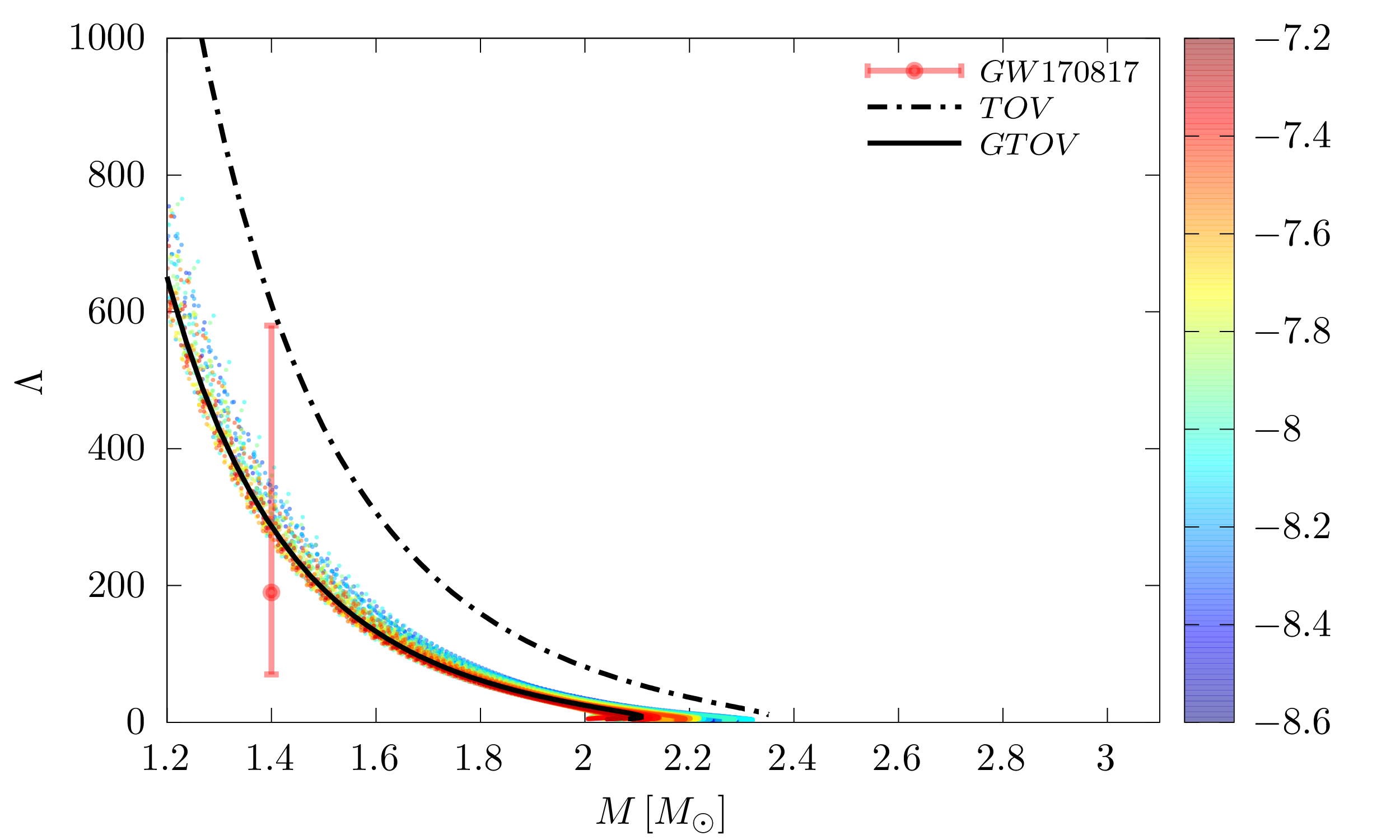}
    \caption{On top we show the corner plot of the posterior distribution of the GTOV parameters, on bottom left we show the mass-radius diagram and on bottom right we show the tidal deformability as a function of the mass. All plots are for Nl3$\omega\rho$Y EoS in the conservative scenario.} \label{nl3wpy_cons}
\end{figure}
%%%%%%%%%%%%%%%%%%%%%%%%%%%%%%%%%%%%%%%%%%%%%%%%%%%%%%%%%%%%%%%%%%%%%%%%%%%%%%%%%%%%

%%%%%%%%%%%%%%%%%%%%%%%%%%%%%%%%%%%%%%%%%%%%%%%%%%%%%%%%%%%%%%%%%%%%%%%%%%%%%%%%%%%%
\begin{figure}[!ht]
    \centering
        \includegraphics[width=0.51\textwidth]{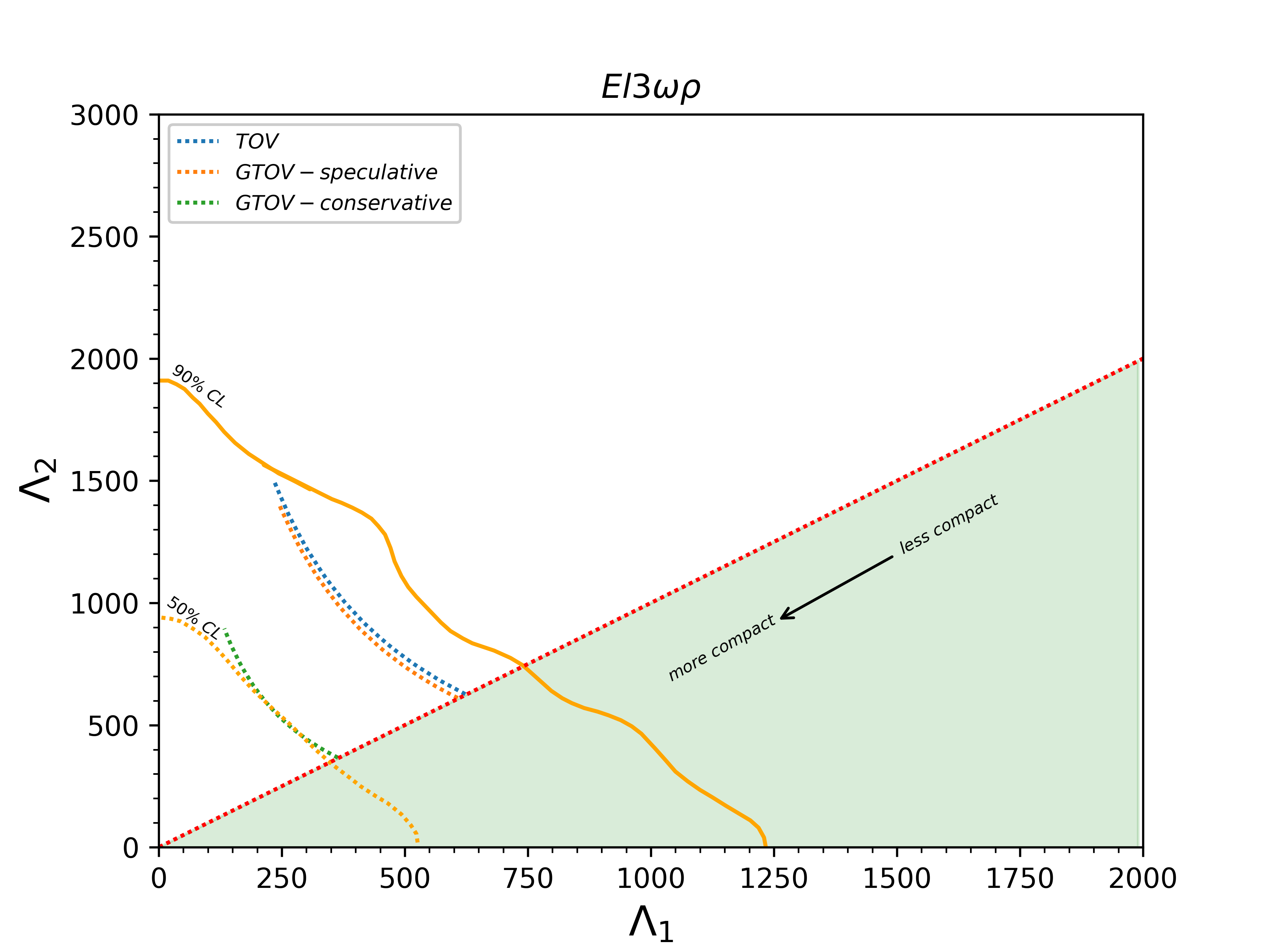}
        \includegraphics[width=0.49\textwidth]{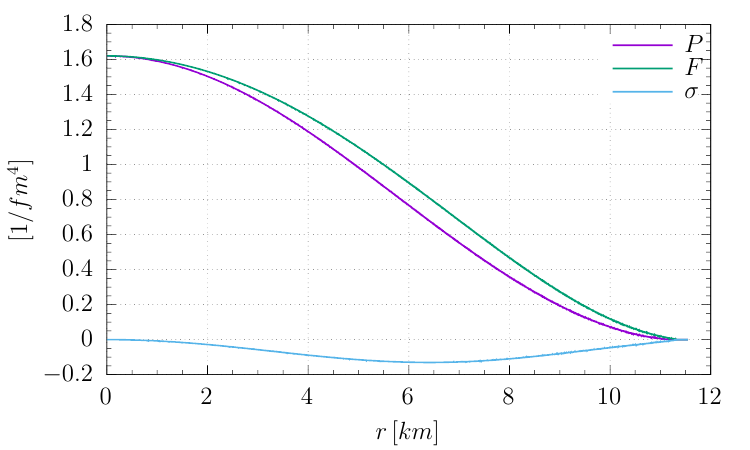}
        \includegraphics[width=0.49\textwidth]{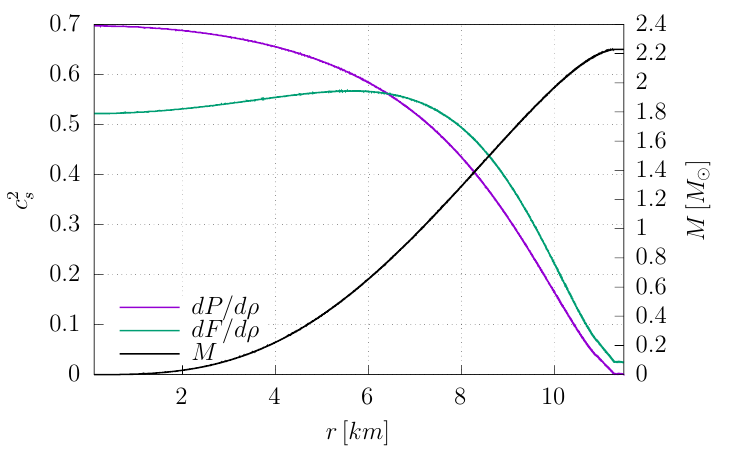}
    \caption{On top we show the dimensionless tidal deformabilities $\Lambda_1$ and $\Lambda_2$ for a binary NS system and the same masses as in GW170817 event. On the bottom left we show the radial $P$ and transverse pressure $F$, and the anisotropy $\sigma$ as functions of the radius $r$. On the bottom right we show the radial $(dP/d\rho)$ and transverse sound velocity $(dF/d\rho)$, and the mass $M$ as functions of the radius $r$. All plots are for a NS with central energy density $\rho_c = 6 \rho_0$ and for El3$\omega\rho$ EoS in the conservative scenario.} \label{el3wp_cons2}
\end{figure}
%%%%%%%%%%%%%%%%%%%%%%%%%%%%%%%%%%%%%%%%%%%%%%%%%%%%%%%%%%%%%%%%%%%%%%%%%%%%%%%%%%%%

%%%%%%%%%%%%%%%%%%%%%%%%%%%%%%%%%%%%%%%%%%%%%%%%%%%%%%%%%%%%%%%%%%%%%%%%%%%%%%%%%%%%
\begin{figure}[!ht]
    \centering
        \includegraphics[width=0.51\textwidth]{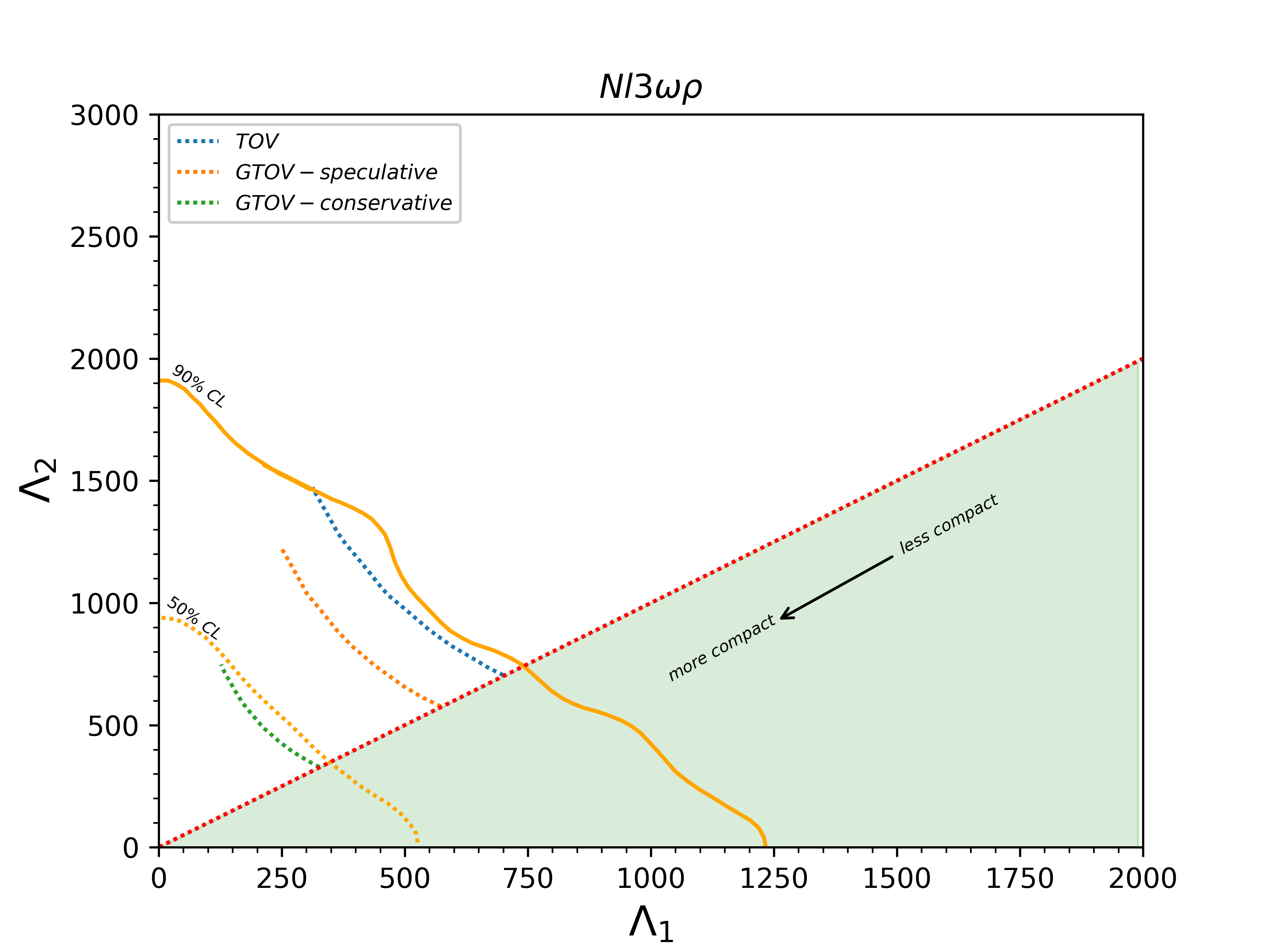}
        \includegraphics[width=0.49\textwidth]{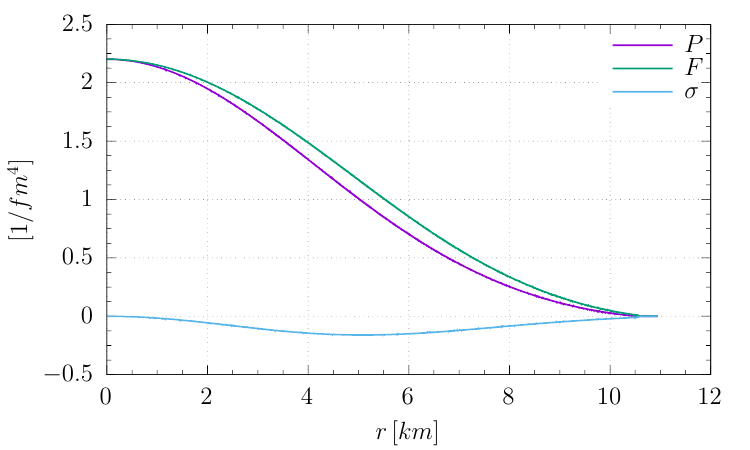}
        \includegraphics[width=0.49\textwidth]{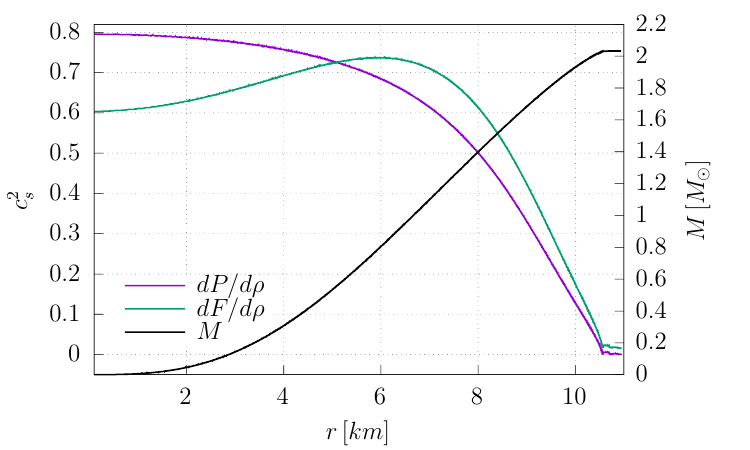}
    \caption{On top we show the dimensionless tidal deformabilities $\Lambda_1$ and $\Lambda_2$ for a binary NS system and the same masses as in GW170817 event. On the bottom left we show the radial $P$ and transverse pressure $F$, and the anisotropy $\sigma$ as functions of the radius $r$. On the bottom right we show the radial $(dP/d\rho)$ and transverse sound velocity $(dF/d\rho)$, and the mass $M$ as functions of the radius $r$. All plots are for a NS with central energy density $\rho_c = 6 \rho_0$ and for Nl3$\omega\rho$ EoS in the conservative scenario.} \label{nl3wp_cons2}
\end{figure}
%%%%%%%%%%%%%%%%%%%%%%%%%%%%%%%%%%%%%%%%%%%%%%%%%%%%%%%%%%%%%%%%%%%%%%%%%%%%%%%%%%%%

%%%%%%%%%%%%%%%%%%%%%%%%%%%%%%%%%%%%%%%%%%%%%%%%%%%%%%%%%%%%%%%%%%%%%%%%%%%%%%%%%%%%
\begin{figure}[!ht]
    \centering
        \includegraphics[width=0.51\textwidth]{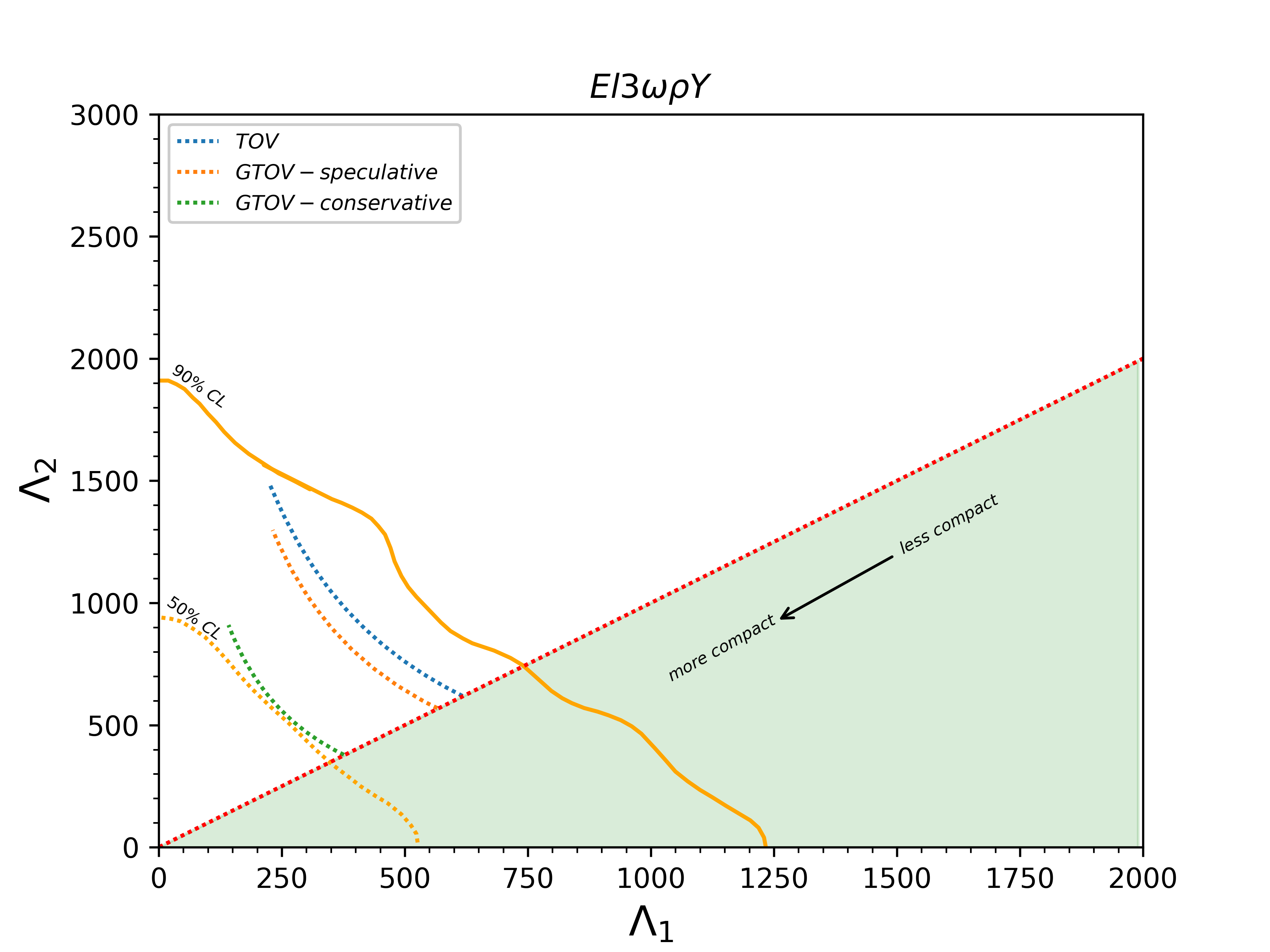}
        \includegraphics[width=0.49\textwidth]{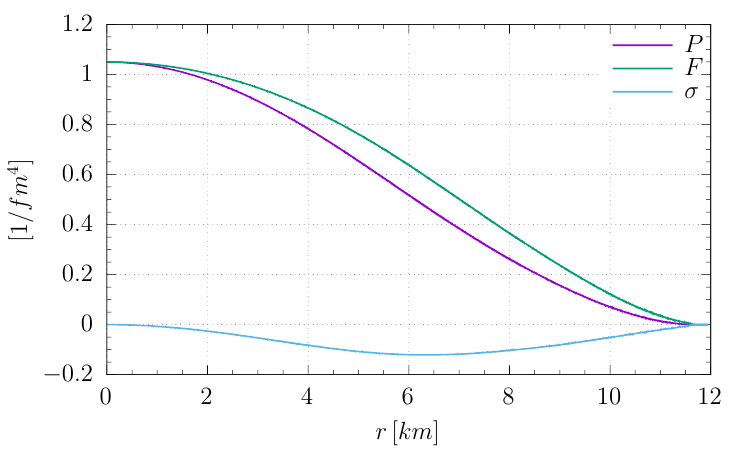}
        \includegraphics[width=0.49\textwidth]{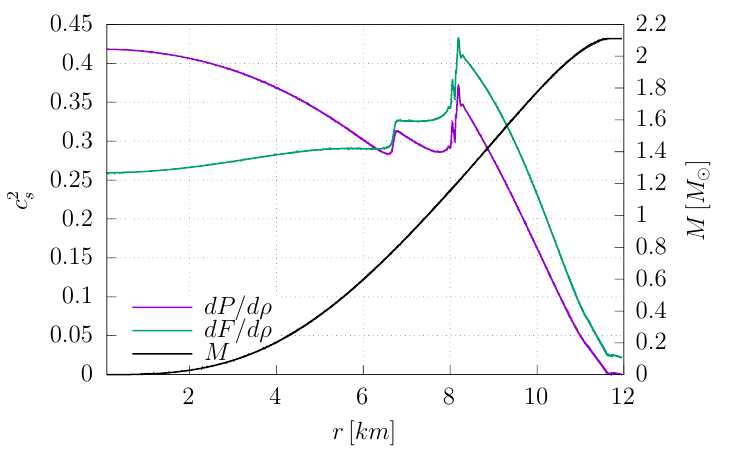}
    \caption{On top we show the dimensionless tidal deformabilities $\Lambda_1$ and $\Lambda_2$ for a binary NS system and the same masses as in GW170817 event. On the bottom left we show the radial $P$ and transverse pressure $F$, and the anisotropy $\sigma$ as functions of the radius $r$. On the bottom right we show the radial $(dP/d\rho)$ and transverse sound velocity $(dF/d\rho)$, and the mass $M$ as functions of the radius $r$. All plots are for a NS with central energy density $\rho_c = 6 \rho_0$ and for El3$\omega\rho$Y EoS in the conservative scenario.} \label{el3wpy_cons2}
\end{figure}
%%%%%%%%%%%%%%%%%%%%%%%%%%%%%%%%%%%%%%%%%%%%%%%%%%%%%%%%%%%%%%%%%%%%%%%%%%%%%%%%%%%%

%%%%%%%%%%%%%%%%%%%%%%%%%%%%%%%%%%%%%%%%%%%%%%%%%%%%%%%%%%%%%%%%%%%%%%%%%%%%%%%%%%%%
\begin{figure}[!ht]
    \centering
        \includegraphics[width=0.51\textwidth]{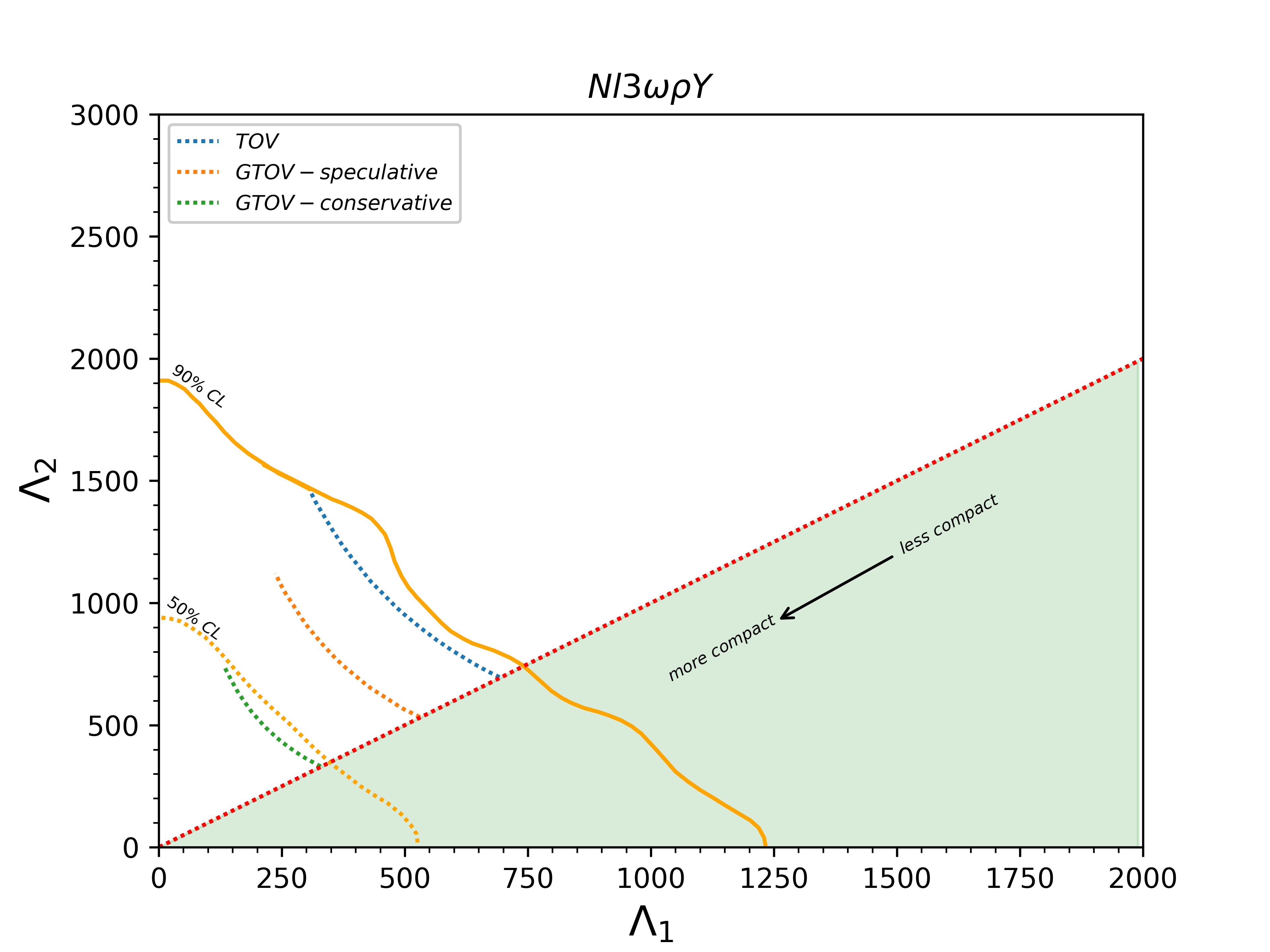}
        \includegraphics[width=0.49\textwidth]{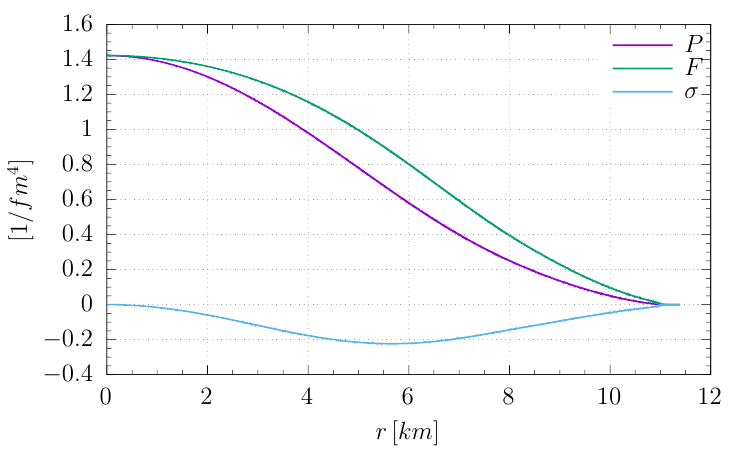}
        \includegraphics[width=0.49\textwidth]{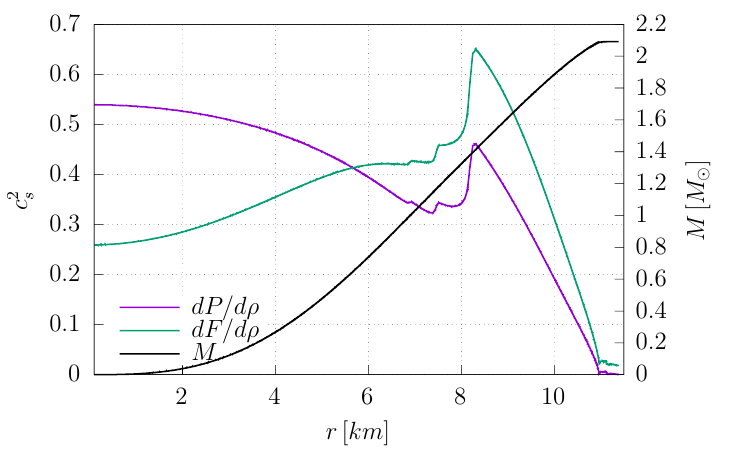}
    \caption{On top we show the dimensionless tidal deformabilities $\Lambda_1$ and $\Lambda_2$ for a binary NS system and the same masses as in GW170817 event. On the bottom left we show the radial $P$ and transverse pressure $F$, and the anisotropy $\sigma$ as functions of the radius $r$. On the bottom right we show the radial $(dP/d\rho)$ and transverse sound velocity $(dF/d\rho)$, and the mass $M$ as functions of the radius $r$. All plots are for a NS with central energy density $\rho_c = 6 \rho_0$ and for Nl3$\omega\rho$Y EoS in the conservative scenario.} \label{nl3wpy_cons2}
\end{figure}
%%%%%%%%%%%%%%%%%%%%%%%%%%%%%%%%%%%%%%%%%%%%%%%%%%%%%%%%%%%%%%%%%%%%%%%%%%%%%%%%%%%%

%%%%%%%%%%%%%%%%%%%%%%%%%%%%%%%%%%%%%%%%%%%%%%%%%%%%%%%%%%%%%%%%%%%%%%%%%%%%%%%%%%%%
\begin{figure}[!ht]
     \centering
             \includegraphics[scale=1.0]{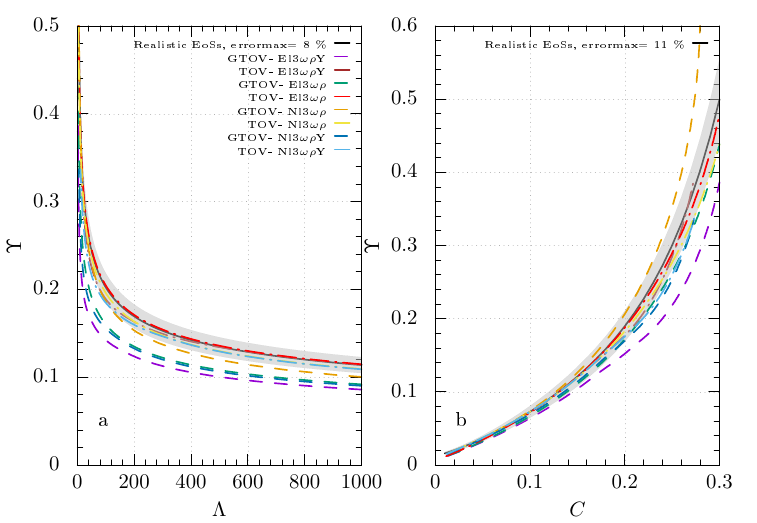}
             \caption{The dashed curves refer to GTOV solutions from each EoS parameterization, the dot-dashed curves to the TOV solution (isotropic case) also from each EoS parameterization, and the solid black line to the fit from Ref.~\cite{Saes:2021fzr}. These results are shown only for the conservative scenario. In the graph on the left, the only GTOV solution that approaches the error band is the one obtained from the parameterization Nl3$\omega\rho$. Meanwhile, in the graph on the right, all the GTOV solutions are within the error range for $0.05 < C < 0.2$, except for the parameterization El3$\omega\rho$, which diverges early.}
\label{upsilonc}
\end{figure}
%%%%%%%%%%%%%%%%%%%%%%%%%%%%%%%%%%%%%%%%%%%%%%%%%%%%%%%%%%%%%%%%%%%%%%%%%%%%%%%%%%%%

In this section, we use Bayesian analysis to optimize the parameters $\alpha$, $\beta$, $\theta$ and $\chi$ from GTOV for the EoSs presented in Sec.~\ref{sec_eos} and assume two possible scenarios. Here we are going to analyze the first scenario, which we have called a conservative scenario, were we try to optimize the GTOV parameters for the mass-radius obtained by NICER for the pulsars PSR J07740$+$6620~\cite{NANOGrav:2019jur} and PSR J0030$+$0451~\cite{riley2019nicer}, the mass-radius of the two NS in GW170817 event~\cite{LIGOScientific:2017vwq,LIGOScientific:2018cki,abbott2019gwtc} and the tidal deformability estimated for this event. In the Appendix \ref{appsec}, we also show the second scenario analyzed, which we called speculative scenario, were we consider the possibility that the compact object within the mass gap in event GW190814~\cite{LIGOScientific:2020zkf} is an NS so we add its mass and tidal deformability as constraints to be fitted.  We assume a uniform prior for the GTOV parameters with the domain of the parameters being taken from the Rahmansyah A., \textit{et al.} paper~\cite{gtovprd}, so that we have: $\alpha \in [-0.15,0.15]$, $\beta \in [0,3]$, $\theta \in [-1,2]$ and $\chi \in [0,1]$. As in~\cite{da2024bayesian}, we assume a Gaussian likelihood function and set causal limits for the mass (M$_{max} < 3.2$ M$_{\odot}$)~\cite{rhoades1974maximum} and, for the radius ($R > 3 G M / c^2 $), and our posterior distributions are obtained using the \textit{emcee} package~\cite{foreman2013emcee}. The package uses the method of Goodman and Weare's Affine Invariant Markov Chain Monte Carlo (MCMC)~\cite{Goodman} for sampling the posterior probability density.

On the top panel of Figs.~\ref{el3wp_cons}, \ref{nl3wp_cons}, \ref{el3wpy_cons} and~\ref{nl3wpy_cons}, we show the corner plots~\cite{corner} of the posterior distributions for the GTOV parameters ($\alpha$, $\beta$, $\theta$, $\chi$) in the conservative scenario when we assume that matter inside the NSs can be described by the El3$\omega\rho$, Nl3$\omega\rho$, El3$\omega\rho$Y and Nl3$\omega\rho$Y EoSs, respectively. The dashed vertical lines in the 1D histograms represent the 0.16, 0.5, and 0.84 quartiles of each histogram. The contour levels in the 2D histograms represent the 0.5, 1, 1.5 and 2$\sigma$ levels containing $11.8\%$, $39.3\%$, $67.5\%$ and $86.4\%$ of the samples for each case, respectively~\cite{corner}.

On the bottom letf panels of Figs.~\ref{el3wp_cons}, \ref{nl3wp_cons}, \ref{el3wpy_cons} and~\ref{nl3wpy_cons}, we show the mass-radius diagrams for the El3$\omega\rho$, Nl3$\omega\rho$, El3$\omega\rho$Y and Nl3$\omega\rho$Y EoSs, respectively, for the conservative scenario. The black dash-dotted curves are the mass-radius for TOV and the black continuous curves are the mass-radius for GTOV using the central values of the parameters obtained in the Bayesian inferences. The colorful dotted curves are the mass-radius curves obtained using values for the GTOV parameters that are within the $68 \%$ Credible Interval (CI) of the log posterior distributions, considering the Highest Density Interval (HDI). The color palette to the right of the plots shows the correspondence between the color of each curve and its log posterior value. These figures also show masses and radii used as constraints in the Bayesian inferences: the two NS in the GW170817 event~\cite{LIGOScientific:2017vwq,LIGOScientific:2018cki,abbott2019gwtc}, one of them with a mass M$_1 = 1.46^{+0.12}_{-0.10}$ M$_{\odot}$ (dark pink continuous horizontal line) and radius R$_1 = 10.8^{+2.0}_{-1.7}$ km (dark pink continuous vertical line) and the other one with a mass of M$_2 = 1.27^{+0.09}_{-0.09}$  M$_{\odot}$ (dark purple continuous horizontal line) and radius R$_2 = 10.7^{+2.1}_{-1.5}$ km (dark purple continuous vertical line); and the two stars from NICER, PSR J07740$+$6620~\cite{NANOGrav:2019jur} with a mass $2.072^{+0.067}_{-0.066}$ M$_{\odot}$ (light pink continuous horizontal line) and radius $12.39^{+1.30}_{-0.98}$ km (light pink continuous vertical line) and PSR J0030$+$0451~\cite{riley2019nicer} with mass $1.34^{+0.15}_{-0.16}$ M$_{\odot}$ (light purple continuous horizontal line) and radius $12.71^{+1.14}_{-1.19}$ km (light purple continuous vertical line); with the shaded regions indicating the uncertainty in the estimated values. 

On the bottom right panels of Figs.~\ref{el3wp_cons}, \ref{nl3wp_cons}, \ref{el3wpy_cons} and~\ref{nl3wpy_cons} we show the dimensionless tidal deformability $\Lambda$ as a function of the mass M [M$_{\odot}$] for the El3$\omega\rho$, Nl3$\omega\rho$, El3$\omega\rho$Y and Nl3$\omega\rho$Y EoSs, respectively, for the conservative scenario. Similarly to the mass-radius diagrams, the black dash-dotted curves represent the GR case and the black continuous curves correspond to the $\Lambda \times$ M [M$_{\odot}$] curves using the central values of the GTOV parameters found in the Bayesian inferences. Here too, the colorful dotted curves are the ones obtained using values from the GTOV parameters that are within the $68 \%$ CI of the log posterior distributions. The value of $\Lambda_{1.4}$ estimated for the GW170817 event, $\Lambda_{1.4} = 190^{+390}_{-120}$ at 90\% confidence level~\cite{LIGOScientific:2017vwq}, is shown as a red dot with its corresponding error bar.
\subsection{El3$\omega\rho$}

From the El3$\omega\rho$ EoS, we observed that the GTOV parameters $\alpha$, $\theta$, and $\chi$ show a significant deviation from the GR scenario in the conservative case as shown on the top panel of Fig.~\ref{el3wp_cons}, particularly, we determined $\theta > 0$. Moreover, the GR model produces a mass-radius diagram that satisfies the NICER and the GW170817 data with $\Lambda_{1.4}$ falling within the upper margin of error of the value determined for the binary system that participated in the GW170817 event, 
as shown on the bottom panels of Fig.~\ref{el3wp_cons}. On the other hand, the GTOV curves have relatively higher radii, but satisfy NICER and the GW170817 data as shown on the same graphs. However, the value of $\Lambda_{1.4}$ falls within the central limit. Generally, $\rm M_{max}$ is reduced and the compactness is higher than in GR. 

\subsection{Nl3$\omega\rho$}

In the case of Nl3$\omega\rho$ EoS, we observe a major deviation of the GTOV parameters from the one obtained from GR. Under this analysis, we found $\theta > 0$ as shown in Fig.~\ref{nl3wp_cons}. With this EoS, we obtain a massive star with $\rm M_{max}= 2.74M_\odot$ in GR as shown in Fig.~\ref{nl3wp_cons}, satisfying the PSR J0030$+$0451 mass-radius data. This lead to $\Lambda_{1.4}$ that strongly satisfies GW190814 and violates GW170817 as presented in Fig.~\ref{nl3wp_cons}. On the other hand, the GTOV curve slightly violates the GW170817 data with comparatively smaller $\rm M_{max}$. This produces a $\Lambda_{1.4}$ that strongly satisfies GW170817. Generally, the GR model produces less compact NSs than the GTOV as the figures show.

\subsection{El3$\omega\rho$Y}

Employing the El3$\omega\rho$Y EoS, which is the addition of hyperons to El3$\omega\rho$ EoS, we highlight the behavior of GTOV parameters relative to the GR model. Analyzing Fig. \ref{el3wpy_cons} we observe significant deviation in the GTOV parameters $\alpha$, $\beta$, and $\chi$, with $\theta > 0$. The addition of hyperons to the El3$\omega\rho$ model reduces the maximum mass achievable from GR, failing to satisfy the mass constraint of PSR J0740$+$6620 and GW170817. However, the GTOV equations satisfy the mass-radius constraint of the NICER pulsars with slightly higher radii (the upper region of the error bar) that do not satisfy the GW data as presented in Fig.~\ref{el3wpy_cons}. The tidal deformability parameter $\Lambda_{1.4}$ increases with hyperons, shifting the $\Lambda_{1.4}$ toward the upper limit of the value determined in the GW170817 event. The mass-radius diagram shows significant differences between GR and GTOV scenarios.

\subsection{Nl3$\omega\rho$Y}

Using the Nl3$\omega\rho$Y EoS which is a hyperonized version of Nl3$\omega\rho$ EoS, we determined the GTOV parameters relative to the GR case once more. For this EoS we also observe a significant deviation of the GTOV parameters from the GR case, with $\theta > 0$. We find that hyperonization of the EoS leads to a reduction in the $\rm M_{max}$ with relatively higher radii which satisfies PSR J0030$+$0451 in the GR case. The GTOV equations, on the other hand, produce a mass-radius relation that is in good agreement with the observed NICER pulsars, however, it fails to satisfy the GW170817 data as displayed in Fig.~\ref{nl3wpy_cons}. Additionally, the hyperonized EoS considered here produces smaller $\Lambda_{1.4}$ in general. However, the GR model mildly satisfies the GW170817 data at its uppermost error limit, on the contrary, the GTOV equations agree well with the GW170817 data as shown in Fig.~\ref{nl3wpy_cons}. 

\section{Further Analysis} \label{sec_plus}

In the left panel of Figs.~\ref{el3wp_cons2}, \ref{nl3wp_cons2}, \ref{el3wpy_cons2} and~\ref{nl3wpy_cons2} we show the dimensionless tidal deformabilities $\Lambda_1$ and $\Lambda_2$ obtained for a binary NS system that has the same chirp mass of the GW170817 event, $\mathcal{M}=1.188~\text{M}_{\odot}$. This way, the masses $m_1$ and $m_2$ of the two NS in the system are related in the following way
\begin{equation}
    \mathcal{M} = \frac{(m_1 m_2)^{3/5}}{(m_1 + m_2)^{1/5}},
\end{equation}
where the values of $m_1$ and $m_2$ are in the same range of values as masses of the NS in GW170817, \textit{i.e.}, $1.36 \leq m_1[\text{M}_{\odot}] \leq 1.58$ and $1.18 \leq m_2[\text{M}_{\odot}] \leq 1.36$. We have calculated $\Lambda_1$ and $\Lambda_2$ for each EoS in GR (blue dotted curve) and in the conservative (green dotted curve) and speculative (orange dotted curve) scenarios using the central values of the GTOV parameters obtained in each inference and using the El3$\omega\rho$, Nl3$\omega\rho$, El3$\omega\rho$Y and Nl3$\omega\rho$Y EoSs. The diagonal red dotted line designates the boundary $\Lambda_1 = \Lambda_2$. We also show the $50\% $ credible levels (orange dotted curve) and $90\%$ CL (orange continuous curve)  determined by LIGO/Virgo in the low-spin prior scenario~\cite{LIGOScientific:2017vwq} for the posteriors obtained using EOS-insensitive relations. 

Based on the afore analysis and Figs.~\ref{el3wp_cons2} and~\ref{nl3wp_cons2}, we can infer that the conservative analysis of GTOV using Nl3$\omega\rho$ yields more compact stars than El3$\omega\rho$, below the 50\% CI (note that the lesser the percentage the more compact the star). The same is true with the speculative case where the curve is between the 90\% CI. Comparing the conservative and the speculative scenarios for the El3$\omega\rho$Y EoS, we observe that the latter results in more compact NSs compared to the former, as shown in Fig.~\ref{el3wpy_cons2}. It is worth mentioning that the stars produced in the speculative scenario using El3$\omega\rho$Y EoS are more compact than the ones built from El3$\omega\rho$.
Lastly, in Fig.~\ref{nl3wpy_cons2}, we observe that the presence of hyperons shows no significant effect on the $\Lambda_1 - \Lambda_2$ diagram in the conservative case. In contrast, stars composed of hyperons are found to be more compact.

On the right plot of Figs.~\ref{el3wp_cons2}, \ref{nl3wp_cons2}, \ref{el3wpy_cons2} and~\ref{nl3wpy_cons2}, the first one shows both radial pressure and transverse pressure against the radius of one star for $\rho=6\rho_{0}$ and $\sigma$, where $\sigma$ shows the behavior between P and F. In the second plot, the transverse and radial speed of sound squared against radius is shown. The NS mass against radius is also presented. The radial speed of sound against $r$, shows the appearance of hyperons through abrupt modifications in its pattern. The hyperons that appear for each drop - for these parameterizations - in the speed of sound  squared are $\Lambda$, $\Sigma^{-}$ and $\Xi^{-}$ (absent only in the plot), respectively. As a requirement for an anisotropic star, we can see -  except for $\rho>0$ and $\frac{d\rho}{dr} <0 $, which is not shown - that all conditions in~\ref{connections} are satisfied.

The only parameters related to observational constraints are $\alpha$ and $\chi$~\cite{Narimani:2014zha, Rappaport:2007ct}. The authors of Ref.~\cite{Narimani:2014zha} extracted $0.04 \lesssim \alpha \lesssim 0.15$ from the observations. The parameter $\chi$ was obtained by fitting the big bang nucleosynthesis (BBN), having the following values: $\chi=1.0 \pm 0.14 $   and  $\chi=0.84 \pm 0.25$. There are two values of  $\chi$ because two values of primordial helium mass fraction ($Y_{p}$) were used~\cite{Rappaport:2007ct}. The fitting of this parameter does not depend strongly on the observational details involved. From conditions (3), (4), and (5) mentioned in~\ref{conditions_ani} near the center of the star $(\sigma_{c}\rightarrow 0)$, the authors of Ref.~\cite{gtovprd} showed that $1/3 \leq \Upsilon_{c} \leq 1$, where $\Upsilon_{c} = P_{c}/\rho_{c}$. The minimum value of $\Upsilon_{c}$ is compatible with that obtained by~\cite{Saes:2021fzr} which is $\Upsilon_{min} \approx 0.3$. 
Universal relations — insensitive~\cite{kent1,kent2} to the EoS — have proven useful for extracting properties of nuclear matter, as they connect macroscopic and microscopic properties of the NS. One way to break the degeneracies in data analysis and model selection for different observations such as radio, X-ray, and gravitational waves is by using universal relations\cite{Yagi:2015hda}. The authors of Ref.~\cite{Yagi:2015hda} found that the I-Love-Q relations are still valid up to 10 $\%$ of the isotropic case. In this work, they used two simple models for pressure anisotropy. In other words, it will depend on the value of the anisotropy used and the model. In our work, this value is associated with the GTOV parameters. Ref.~\cite{Saes:2021fzr} found for GR an approximate universal relation between $\Upsilon - C/\Lambda/\bar{I}$ for 25 realistic EoSs. The  $\Upsilon$ can be viewed as a mean notion of the stiffness of nuclear matter inside the NS. The tolerance of universal relations varies for each pair. Therefore, the maximum error is different for each graph below. The same occurs in~\cite{Yagi:2015hda} for I-Love-Q relations. 

In Fig.~\ref{upsilonc}, one can note that $\Upsilon - C/\Lambda$ relations are maintained for isotropic case GR. Meanwhile, for the case of anisotropic case (GTOV), only $\Upsilon - C$ are within the error band for certain EoSs and GTOV parameters. It is worth mentioning that for other values of $\alpha$, $\beta$, $\theta$, and $\chi$, it would be possible to be within the error band provided by Ref.~\cite{Saes:2021fzr}, \textit{i.e.}, for a weaker anisotropy. If the calculated relations fall within the error band, it is possible to estimate the dimensionless moment of inertia with relatively good precision.

%%%%%%%%%%%%%%%%%%%%%%%%%%%%%%%%%%%%%%%%%%%%%%%%%%%%%%%%%%%%%%%%%%%%%%%%%%%%%%%%%%%%

\vspace{0.5cm}
\begin{table}[!ht]
\centering
\begin{tabular}{ |p{2cm}|p{2cm}|p{4cm}|p{2cm}|p{2cm}|p{2cm}|p{2cm}|  }
 \hline
 EoS & Scenario  & $(\alpha,\beta,\theta, \chi)$  & $M_{max}$ $[M_{\odot}]$ & $R_{max}$ [km] & $\Lambda_{1.4}$ & $R_{1.4}$ [km]  \\ 
 \hline
 \hline
 El3$\omega\rho$Y&   speculative   & (-0.03, 0.77,-0.78, 0.36)&      2.57 \textcolor{cyan}{ \checkmark} &  11.72    &  490 \textcolor{red}{ \checkmark} \textcolor{blue}{ \checkmark}   & 13.44   \textcolor{violet}{\checkmark}  \textcolor{BurntOrange}{\checkmark}\\  \hline
 El3$\omega\rho$ &   speculative   &(-0.02, 1.85, -0.44, 0.57)&      2.53 \textcolor{cyan}{ \checkmark} &  11.87    &  509 \textcolor{red}{ \checkmark} \textcolor{blue}{ \checkmark} & 13.10  \textcolor{violet}{\checkmark}  \textcolor{BurntOrange}{\checkmark} \\  \hline
 Nl3$\omega\rho$Y &  speculative   &(-0.02, 1.13, -0.08, 0.44)&      2.63 \textcolor{cyan}{ \checkmark} &  13.18    &  467 \textcolor{red}{ \checkmark} \textcolor{blue}{ \checkmark} & 13.62  \textcolor{violet}{\checkmark}  \textcolor{BurntOrange}{\checkmark}\\  \hline
 Nl3$\omega\rho$ &   speculative   &(-0.01, 1.85, 0.24, 0.63)&      2.62 \textcolor{cyan}{ \checkmark} &  12.49    &  504 \textcolor{red}{ \checkmark} \textcolor{blue}{ \checkmark} & 13.40  \textcolor{violet}{\checkmark}  \textcolor{BurntOrange}{\checkmark} \\  \hline
 El3$\omega\rho$Y&   conservative   &(-0.08, 1.21, 0.38, 0.53)&      2.12 \textcolor{Sepia}{ \checkmark} &  11.37 \textcolor{BlueViolet}{\checkmark}    &  323 \textcolor{red}{ \checkmark}  & 13.14  \textcolor{violet}{\checkmark} \textcolor{BurntOrange}{\checkmark}  \\  \hline
 El3$\omega\rho$ &   conservative   &(-0.07, 1.74, 0.52, 0.62)&      2.24 \textcolor{Sepia}{ \checkmark}&  11.31 \textcolor{BlueViolet}{\checkmark}   &  312 \textcolor{red}{ \checkmark}& 12.94  \textcolor{violet}{\checkmark} \textcolor{BurntOrange}{\checkmark}  \\  \hline
 Nl3$\omega\rho$Y &  conservative   &(-0.06, 1.66, 1.57, 0.58)&      2.19 \textcolor{Sepia}{ \checkmark} &  11.74 \textcolor{BlueViolet}{\checkmark}   &  287 \textcolor{red}{ \checkmark} & 13.13  \textcolor{violet}{\checkmark}\textcolor{BurntOrange}{\checkmark}  \\  \hline
 Nl3$\omega\rho$ &   conservative  &(-0.05, 2.5, 1.71, 0.75)&      2.06 \textcolor{Sepia}{ \checkmark} &  11.71 \textcolor{BlueViolet}{ \checkmark}   &  282 \textcolor{red}{ \checkmark} & 12.91   \textcolor{violet}{\checkmark} \textcolor{BurntOrange}{\checkmark}\\  \hline
 El3$\omega\rho$Y&  GR   &(0,  1, 0, 1)&      1.96  &  11.43    &  530 \textcolor{red}{ \checkmark} \textcolor{blue}{ \checkmark} & 12.81   \textcolor{violet}{\checkmark} \textcolor{BurntOrange}{\checkmark} \\  \hline
 El3$\omega\rho$ &  GR   &(0, 1, 0, 1)&      2.30 \textcolor{Sepia}{ \checkmark} &  11.22 \textcolor{BlueViolet}{\checkmark}   &  536 \textcolor{red}{ \checkmark} \textcolor{blue}{ \checkmark}& 12.81  \textcolor{violet}{\checkmark} \textcolor{BurntOrange}{\checkmark} \\  \hline
 Nl3$\omega\rho$Y & GR   &(0, 1, 0, 1)&      2.35 \textcolor{Sepia}{ \checkmark}   &  12.73 \textcolor{BlueViolet}{\checkmark}   &  611 \textcolor{blue}{ \checkmark} & 13.45 \textcolor{violet}{\checkmark}  \textcolor{BurntOrange}{\checkmark}  \\  \hline
 Nl3$\omega\rho$ &  GR   &(0, 1, 0, 1)&      2.74 \textcolor{Sepia}{ \checkmark}  &  12.64   &  619 \textcolor{blue}{ \checkmark}& 13.45 \textcolor{violet}{\checkmark}  \textcolor{BurntOrange}{\checkmark} \\  \hline
\end{tabular}
\caption{Here, we make a comparison between our results and the values used in the Bayesian analysis. For the dimensionless tidal deformability, we use \textcolor{red}{$\Lambda_{1.4} (GW170817) = 190^{+390}_{-120}$} and  \textcolor{blue}{$\Lambda_{1.4} (GW190814) = 616^{+273}_{-158}$}. While for masses and radii, we have:  \textcolor{cyan}{mass (GW190814) = $2.5^{+0.08}_{-0.09}$ M$_{\odot}$}, \textcolor{Sepia}{ mass(PSR J07740$+$6620) = $2.072^{+0.067}_{-0.066}$ M$_{\odot}$} and \textcolor{BlueViolet}{ radius $12.39^{+1.30}_{-0.98}$ km}, \textcolor{BurntOrange}{ mass (PSR J0030$+$0451) = $1.34^{+0.15}_{-0.16}$ M$_{\odot}$} and \textcolor{violet}{ radius $12.71^{+1.14}_{-1.19}$ km}}. 
\label{table_values}
\end{table}

%%%%%%%%%%%%%%%%%%%%%%%%%%%%%%%%%%%%%%%%%%%%%%%%%%%%%%%%%%%%%%%%%%%%%%%%%%%%%%%%%%%%

\section{Conclusions} \label{conclusions}

In this work, we use Bayesian inference to optimize the GTOV parameters to satisfy recent astrophysical data from NICER and GW detections, considering four EoS parameterizations. We set our prior distributions based on the values for the GTOV parameters obtained in~\cite{gtovprd}. Unlike in their work, we do not define each parameter set according to a physical condition such as $\sigma=0$ at the center of the star, but rather use the minimum and maximum values of each parameter suggested by them. Using the available observational data about mass, radius, and dimensionless tidal deformability for the likelihood distribution, we find the condition $\sigma=0$ at the center of the star, as can be seen on the bottom left panels of Figs.~\ref{el3wp_cons2}, \ref{nl3wp_cons2}, \ref{el3wpy_cons2} and~\ref{nl3wpy_cons2}. However, using their expression for $\sigma$, we obtained a negative alpha. Moreover, if we analyze only the mass-radius diagram without calculating the dimensionless tidal deformability in our code, the alpha can be positive in agreement with~\cite{clesio,gtovprd}. However, when including the dimensionless tidal deformability, the code runs only for negative alpha. This is mainly because when computing the dimensionless tidal deformability the value of the denominator $( 1 - d \sigma / dP )$  tends to zero\footnote{If this term goes to zero, it means that there is no physical solution at the center of the star. Because $( 1 - d \sigma / dP )$  $= dF/dP=c_s^2 (transverse)/c_s^2 (radial) $ and, from anisotropic conditions, both speed of sound must be greater than 0 and less than 1.} for $\alpha>0$, leading to unphysical solutions.  In~\cite{velten2016}, negative values of $\alpha$ were also tested, and the authors concluded that values of alpha smaller than zero lead to an increase in the mass and the radius. We must point out that in their work, the dimensionless tidal deformability was not calculated. In our work, however, we found that when we combine the variation of the other three parameters ($\beta,~\theta$ and $\chi$) simultaneously it is possible to have $\alpha<0$ and an increase on both the mass and the radius, see Fig.~\ref{nl3wp_cons} for example. 

We used two possible constraint scenarios - conservative and speculative - and four different EoS parameterizations to constrain the GTOV parameters. In the conservative scenario, we used mass-radius data from NICER and the GW170817 event plus the data for the tidal deformability from the GW170817, and in the speculative scenario, we also considered the possibility that the compact object that falls within the mass gap in the GW190814 event is a NS, so its mass and deformability were added to the constraints. In Tab.~\ref{table_values}, one can notice how the GTOV parameters changed $M,~R$, and $\Lambda$ to be close to the observational data. We can observe that with GTOV we were able to satisfy the constraints of tidal deformability coming from the GW events in all cases considered. However, in the mass-radius diagram and tidal deformability, we see that none of these results satisfied all the imposed constraints at the same time. Another interesting aspect is that using GTOV both parameterizations, with and without hyperons, lead to similar values for $M,~R$, and  $\Lambda$. Besides, we emphasize that without the parameterization of the $\phi$ field related to EoS with hyperons (if $\alpha_v < 1$, this cause an increase in mass-radius relation, here we used $\alpha_v =1$) the degree of anisotropy from GTOV would be higher since it would be necessary to increase the pressure to achieve an NS with a mass $\sim$ $2M_{\odot}$
 
On the other hand, when we disregard the masses and radii of $M_{1}$ and $M_{2}$ from GW170817, we see that both the TOV and the GTOV present better results in both the conservative and the speculative scenario. In the conservative scenario, all results approach the central value of the tidal deformability from GW170817. Three of our results are within the expected $\chi$ value set by~\cite{Rappaport:2007ct}: Nl3$\omega\rho$ in both the conservative and the speculative scenario and El3$\omega\rho$ in the conservative scenario. As for the $\alpha$ values, we cannot make a direct comparison with previous results, since our results have a different $\alpha$-sign. 

Additionally, we analyzed the approximate universal relations of $\Upsilon - C/\Lambda$. In GR, all the parameterizations used here respected the error margin. However, for GTOV the relation $\Upsilon - C/\Lambda$ gives a good result only for Nl3$\omega\rho$ in the conservative scenario, since the values of $\Lambda < 200$ and $C < 0.25$ of these relations are satisfied. Some works have recalculated these universal relations for anisotropic stars with realistic EoS~\cite{Biswas:2019gkw,Das:2022ell}. However, assuming that the results from modified theories of gravity should not diverge significantly from those of GR, the same applies here. In other words, the universal relations obtained for isotropic stars using realistic EoSs are essential for determining the error margin in our results for GTOV. It is worth noticing that in the GTOV context, the anisotropy arises from modified theories of gravity and not from an anisotropy model~\cite{Bowers,Horvat_2011,Cosenza:1981myi}.

From our analysis, we can conclude that the conservative scenario with the Nl3$\omega\rho$ EoS provides the best agreement with the constraints, so that, our best estimate for the values of the GTOV parameters is the one found for this case, that is, $\alpha = -0.05 \pm 0.03$, $\beta = 2.50^{+0.30}_{-0.41}$, $\theta = 1.71^{+0.18}_{-0.29}$ and $\chi = 0.75^{+0.15}_{-0.20}$. By using the GTOV formalism and the Nl3$\omega\rho$ EoS, we can simultaneously satisfy the mass-radius constraints from NICER and LIGO and the tidal deformability constraints from GW170817, with only a small deviation in the values of the parameters $\alpha$ and $\chi$ when compared to GR. Additionally, the value of $\chi$ in this case is similar to the one encountered in~\cite{Rappaport:2007ct}. On the other hand, the parameters $\beta$ and $\theta$ present a more significant deviation from GR. We can conclude that this more significant deviation primarily arises from the need to obtain a value of $\Lambda_{1.4}$ that aligns with the central value encountered in~\cite{LIGOScientific:2017vwq}. We emphasize that different EoS parameterizations can lead to different values for the GTOV parameters so that, an EoS that aligns better with the constraints is expected to reduce intervals of the deviations in the GTOV parameters. 
Additionally, since the GTOV formalism can also be associated to anisotropy, as demonstrated in~\cite{gtovprd}, further research on the potential sources of anisotropy inside the NSs -- such as extremely strong magnetic fields, phase transitions, pion condensation, core crystallization, relativistic nuclear interactions, and superfluid cores (for a review on anisotropy, see~\cite{Herrera:1997plx}) -- could help us to have a better understanding of the physical meaning of the deviations on the values of the GTOV parameters.

\section*{Acknowledgements}
We would like to thank A. Sulaksono and A. Rahmansyah for the helpful feedback on their calculations. We thank Luiz L. Lopes for useful discussions and suggestions.
This work is a part of the project INCT-FNA proc. No. 464898/2014-5. It is also supported by Conselho Nacional de Desenvolvimento Cient\'ifico e Tecnol\'ogico (CNPq) under Grant No. 303490/2021-7 (D.P.M.). F.K and L.C.N.S would like to thank FAPESC/CNPq for financial support under grants 174332/2023-8 and 735/2024. A.I. would like to thank the S\~ao Paulo State Research Foundation (FAPESP) for financial support through Grant No.  2023/09545-1.

\bibliographystyle{ieeetr}
\bibliography{ref.bib}

\appendix

\section{Speculative Scenario} \label{appsec}

%%%%%%%%%%%%%%%%%%%%%%%%%%%%%%%%%%%%%%%%%%%%%%%%%%%%%%%%%%%%%%%%%%%%%%%%%%%%%%%%%%%%
\begin{figure}[!ht] 
    \centering
        \includegraphics[width=0.51\textwidth]{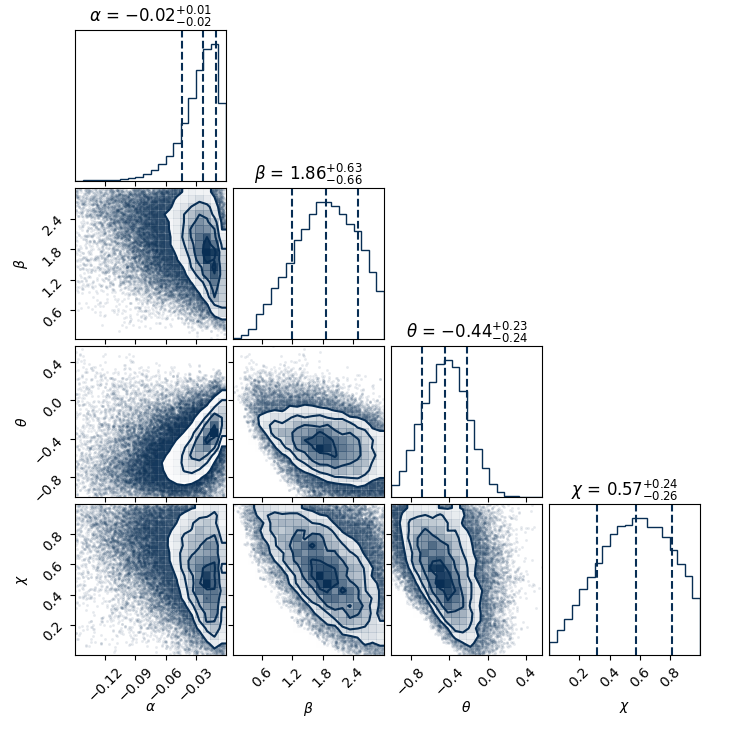}
        \includegraphics[width=0.49\textwidth]{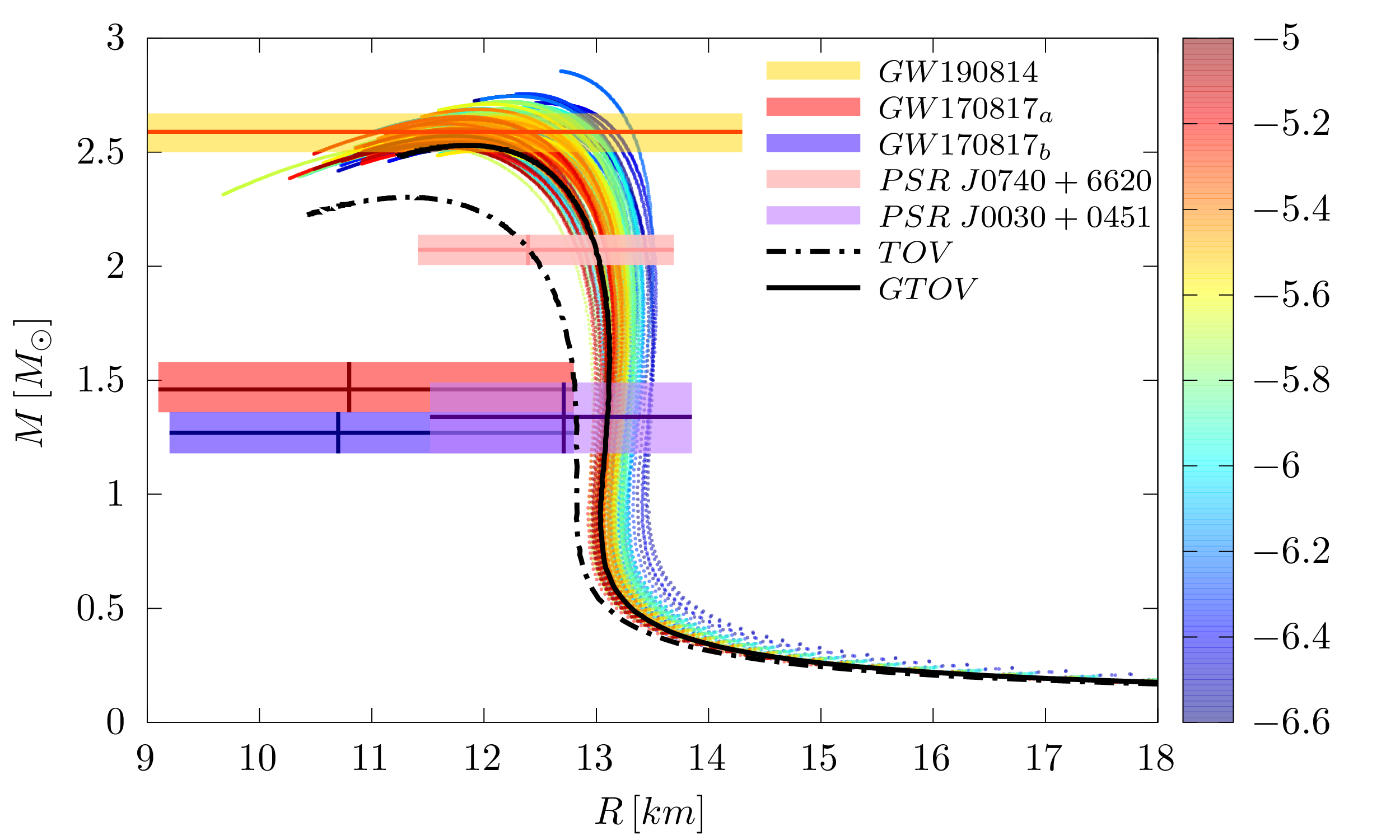}
        \includegraphics[width=0.49\textwidth]{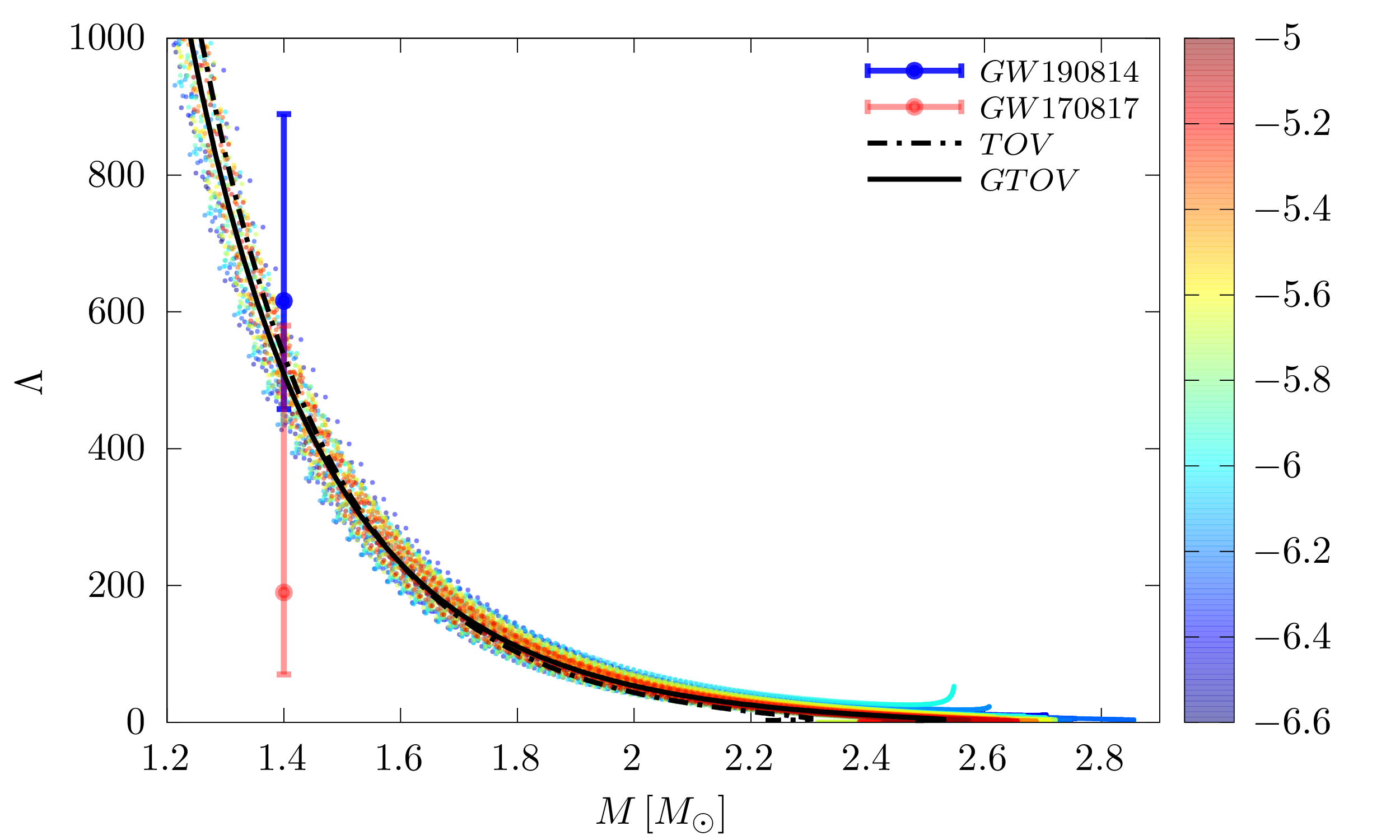}
    \caption{On top we show the corner plot of the posterior distribution of the GTOV parameters, on bottom left we show the mass-radius diagram and on bottom right we show the tidal deformability as a function of the mass. All plots are for El3$\omega\rho$ EoS in the speculative scenario.} \label{el3wp_spec}
\end{figure}
%%%%%%%%%%%%%%%%%%%%%%%%%%%%%%%%%%%%%%%%%%%%%%%%%%%%%%%%%%%%%%%%%%%%%%%%%%%%%%%%%%%%

%%%%%%%%%%%%%%%%%%%%%%%%%%%%%%%%%%%%%%%%%%%%%%%%%%%%%%%%%%%%%%%%%%%%%%%%%%%%%%%%%%%%
\begin{figure}[!ht] 
    \centering
        \includegraphics[width=0.51\textwidth]{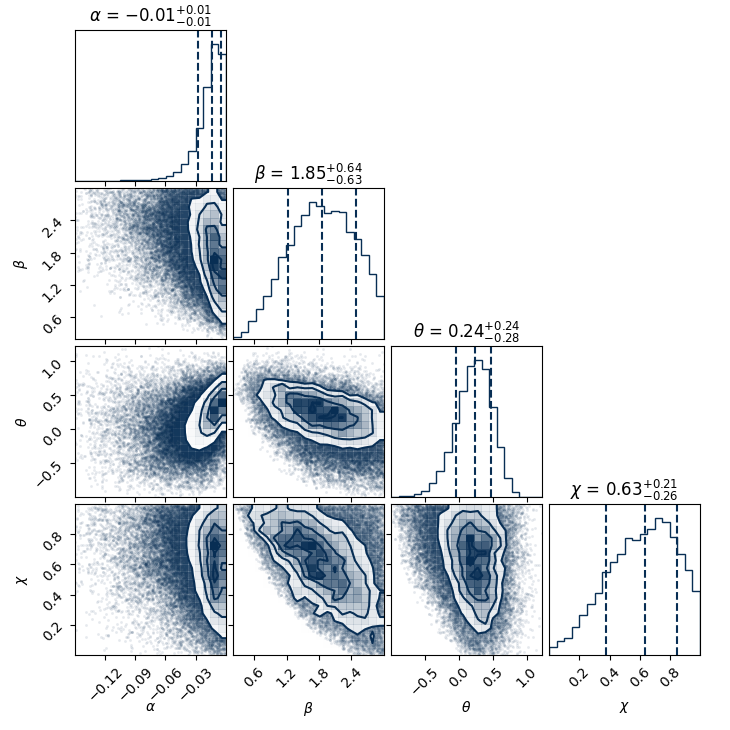}
        \includegraphics[width=0.49\textwidth]{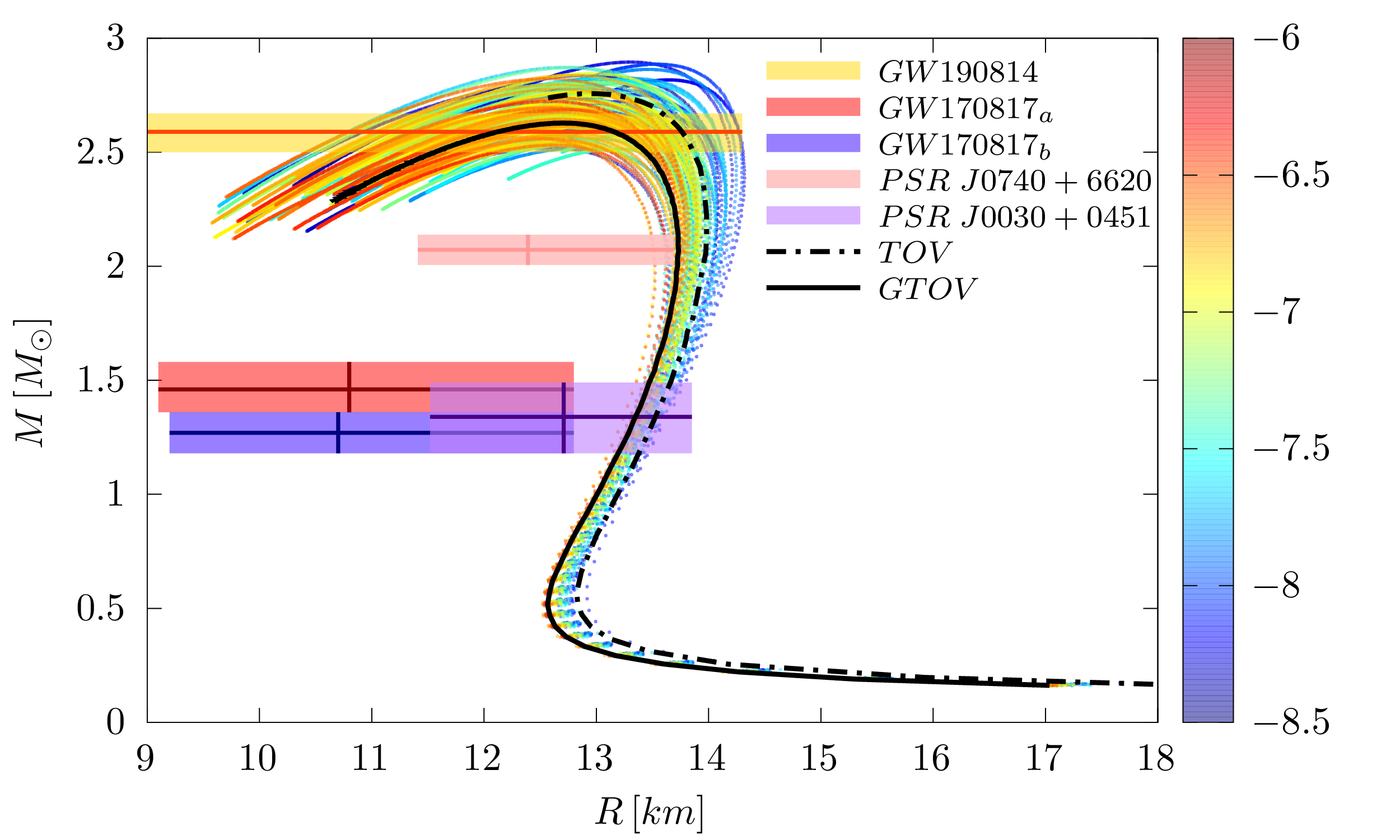}
        \includegraphics[width=0.49\textwidth]{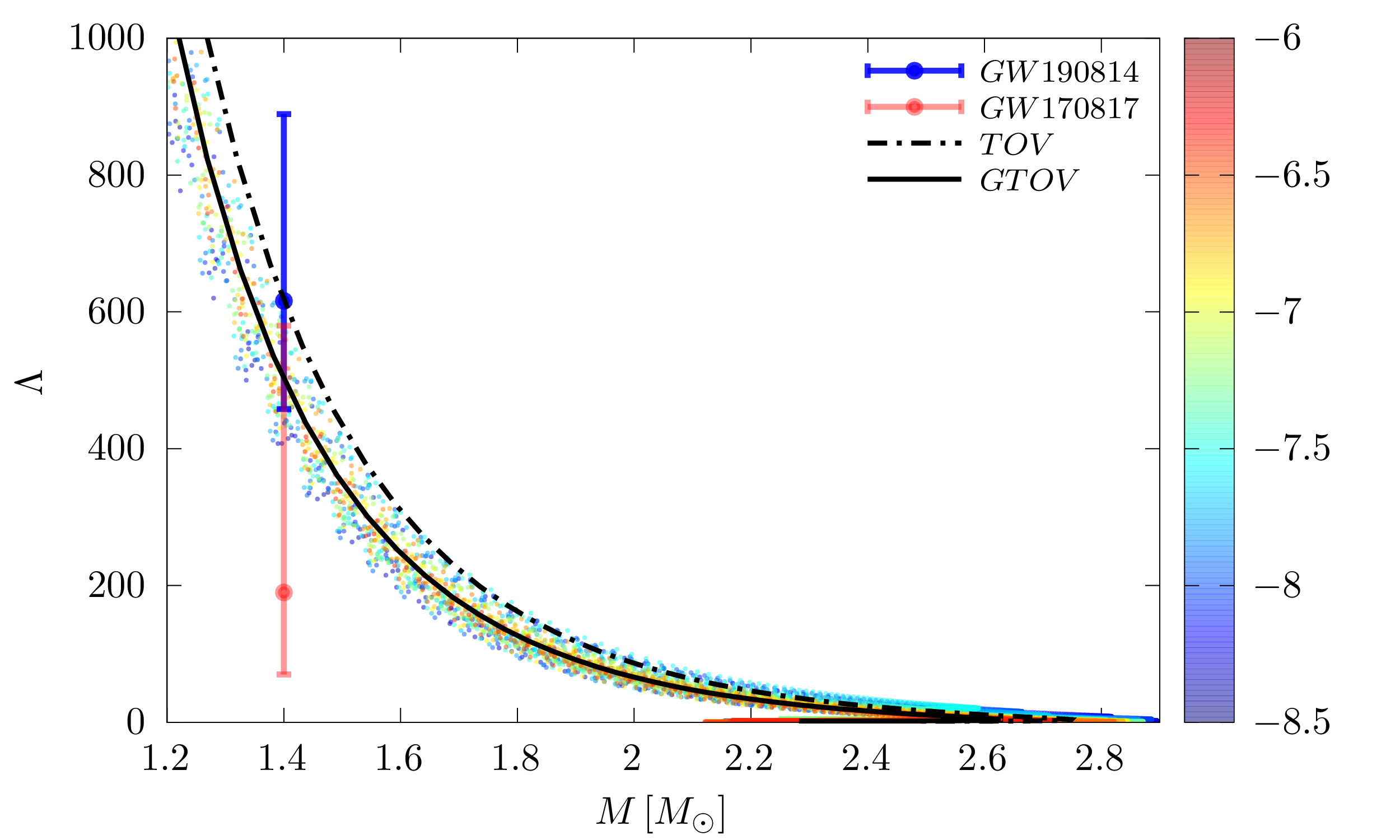}
    \caption{On top we show the corner plot of the posterior distribution of the GTOV parameters, on bottom left we show the mass-radius diagram and on bottom right we show the tidal deformability as a function of the mass. All plots are for Nl3$\omega\rho$ EoS in the speculative scenario.} \label{nl3wp_spec}
\end{figure} 
%%%%%%%%%%%%%%%%%%%%%%%%%%%%%%%%%%%%%%%%%%%%%%%%%%%%%%%%%%%%%%%%%%%%%%%%%%%%%%%%%%%%

%%%%%%%%%%%%%%%%%%%%%%%%%%%%%%%%%%%%%%%%%%%%%%%%%%%%%%%%%%%%%%%%%%%%%%%%%%%%%%%%%%%%
\begin{figure}[!ht]
    \centering
        \includegraphics[width=0.51\textwidth]{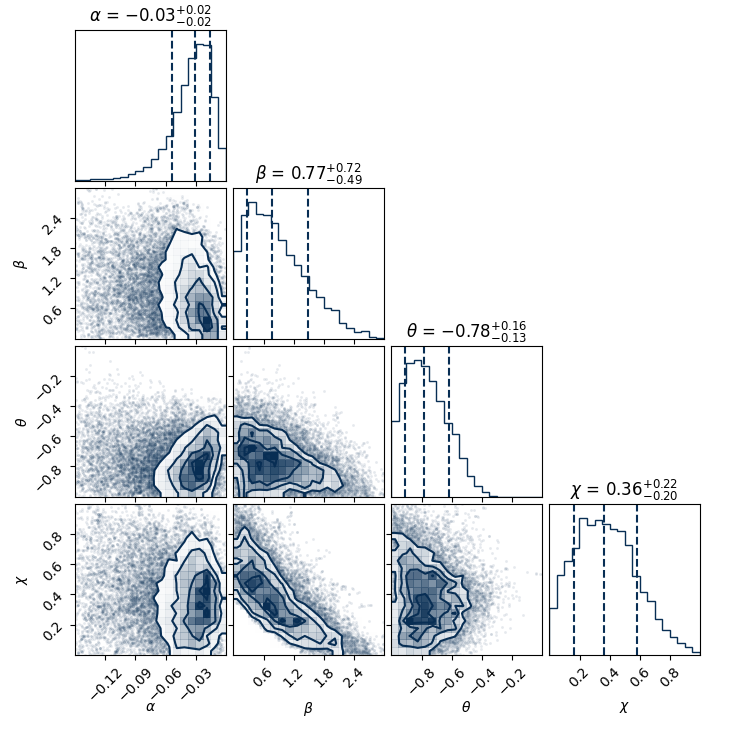}
        \includegraphics[width=0.49\textwidth]{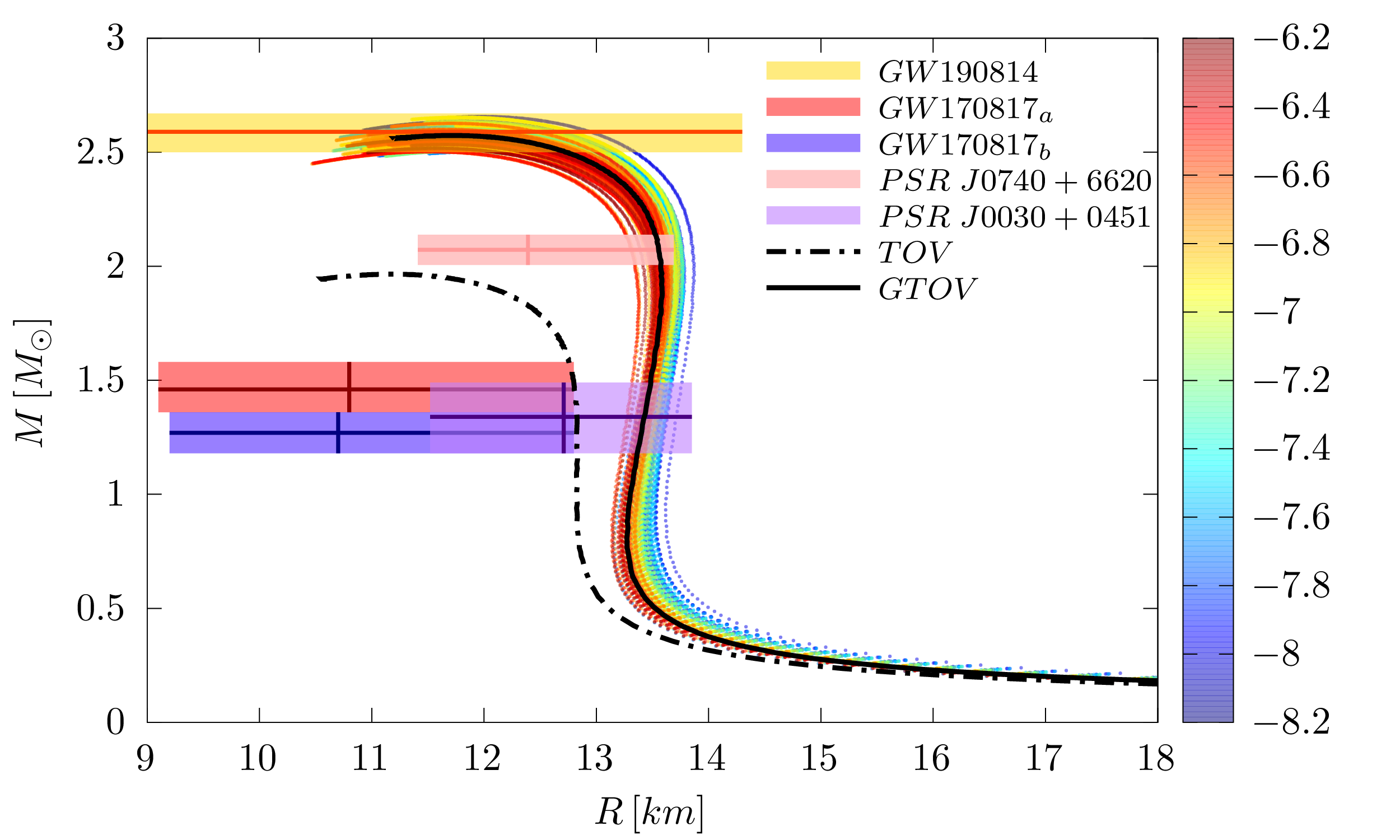}
        \includegraphics[width=0.49\textwidth]{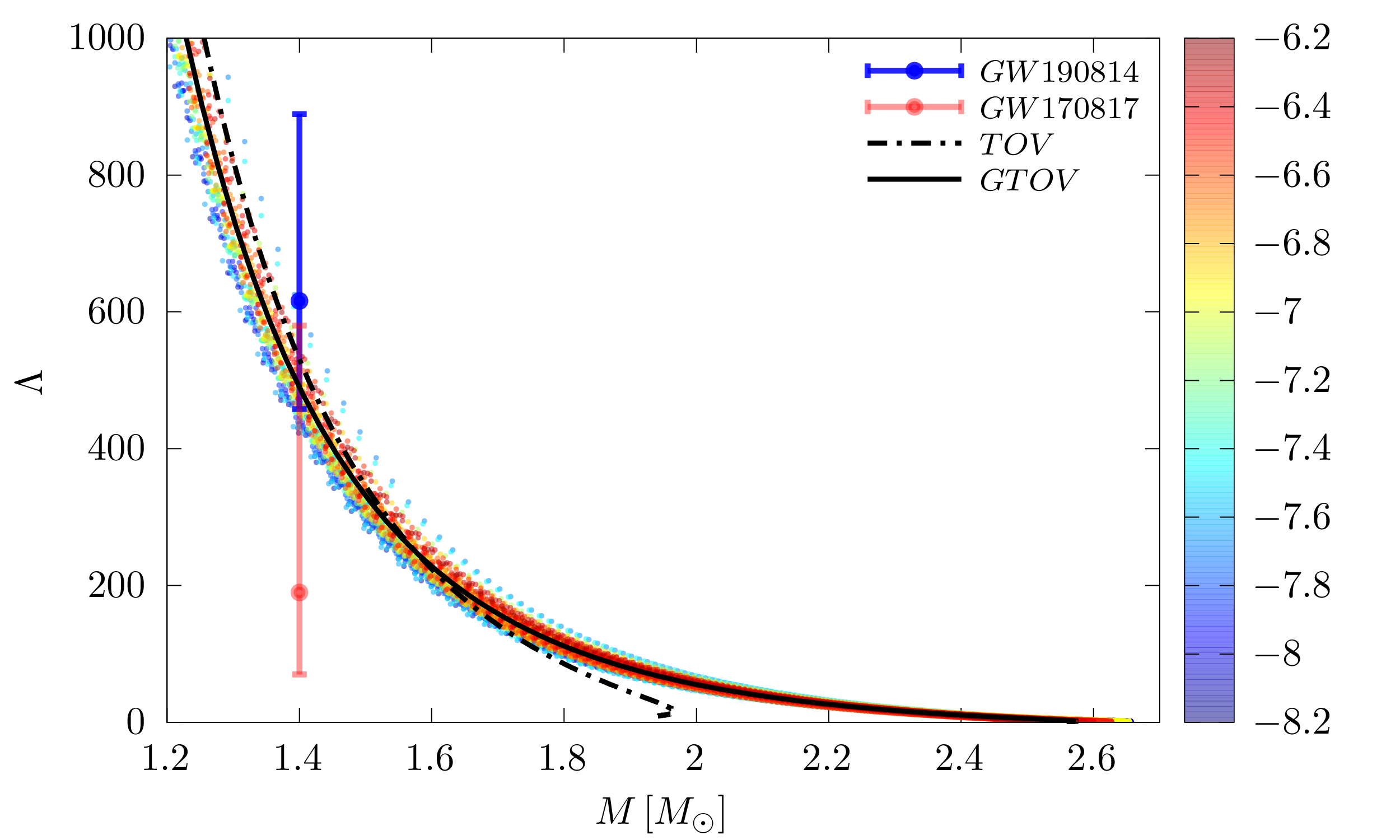}
    \caption{On top we show the corner plot of the posterior distribution of the GTOV parameters, on bottom left we show the mass-radius diagram and on bottom right we show the tidal deformability as a function of the mass. All plots are for El3$\omega\rho$Y EoS in the speculative scenario.} \label{el3wpy_spec}
\end{figure}
%%%%%%%%%%%%%%%%%%%%%%%%%%%%%%%%%%%%%%%%%%%%%%%%%%%%%%%%%%%%%%%%%%%%%%%%%%%%%%%%%%%%

%%%%%%%%%%%%%%%%%%%%%%%%%%%%%%%%%%%%%%%%%%%%%%%%%%%%%%%%%%%%%%%%%%%%%%%%%%%%%%%%%%%%
\begin{figure}[!ht]
    \centering
        \includegraphics[width=0.51\textwidth]{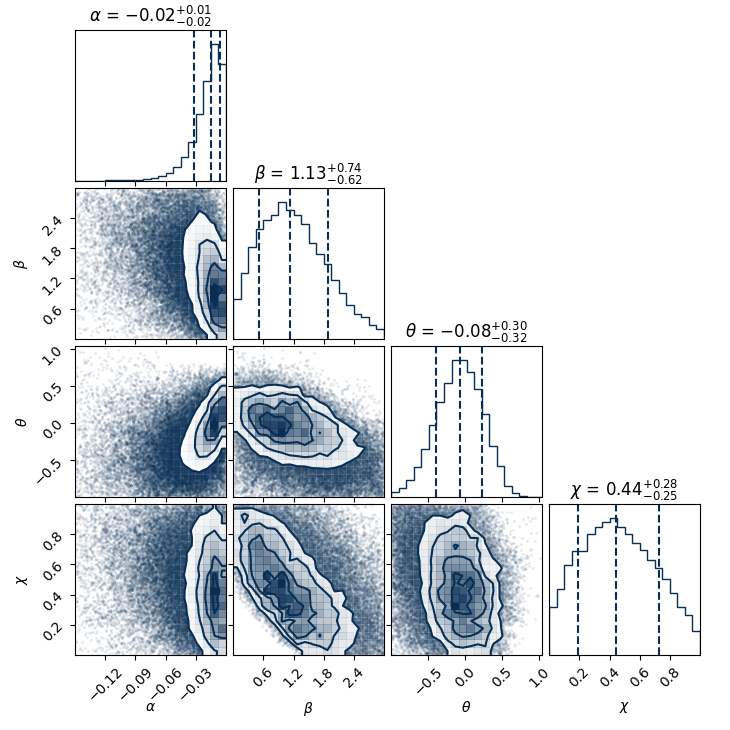}
        \includegraphics[width=0.49\textwidth]{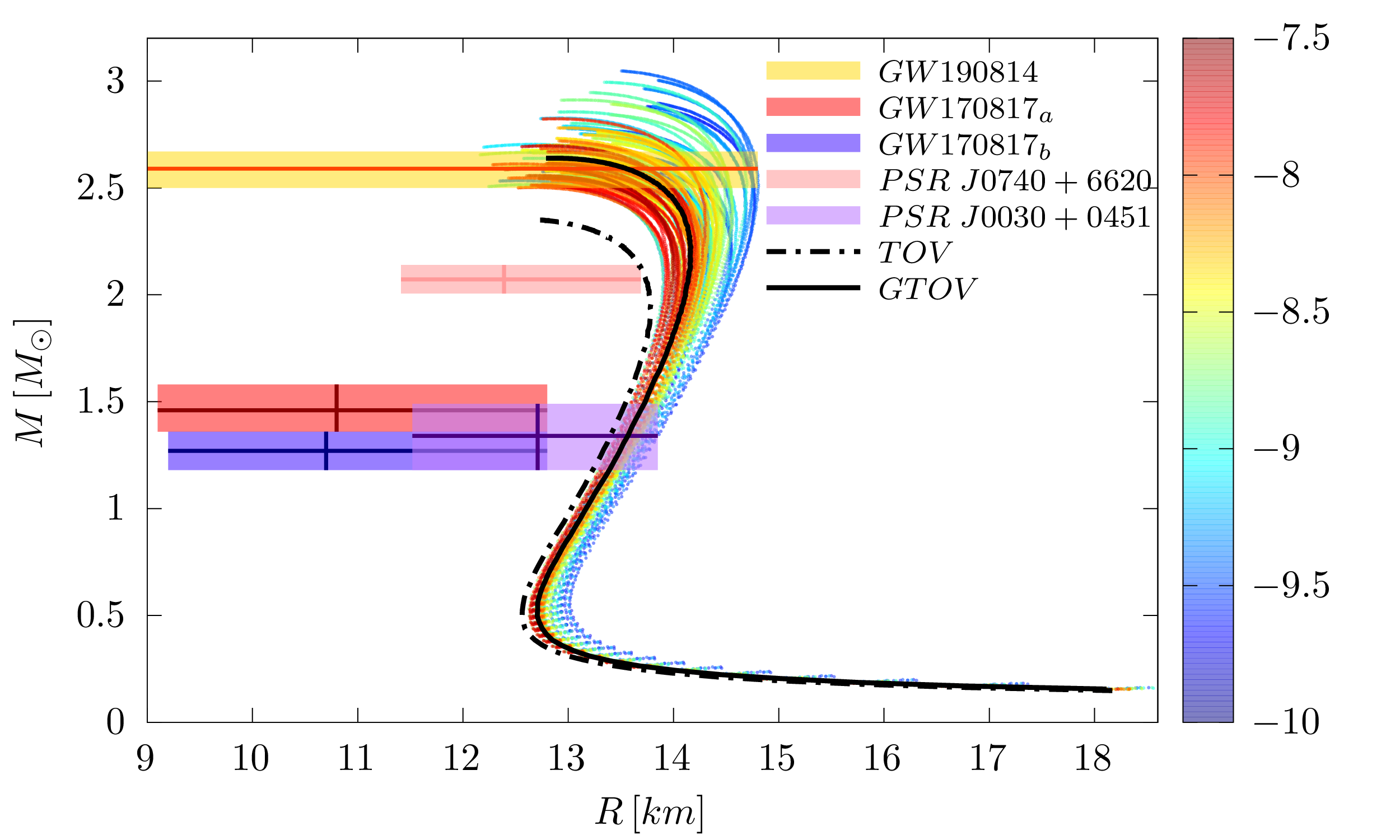}
        \includegraphics[width=0.49\textwidth]{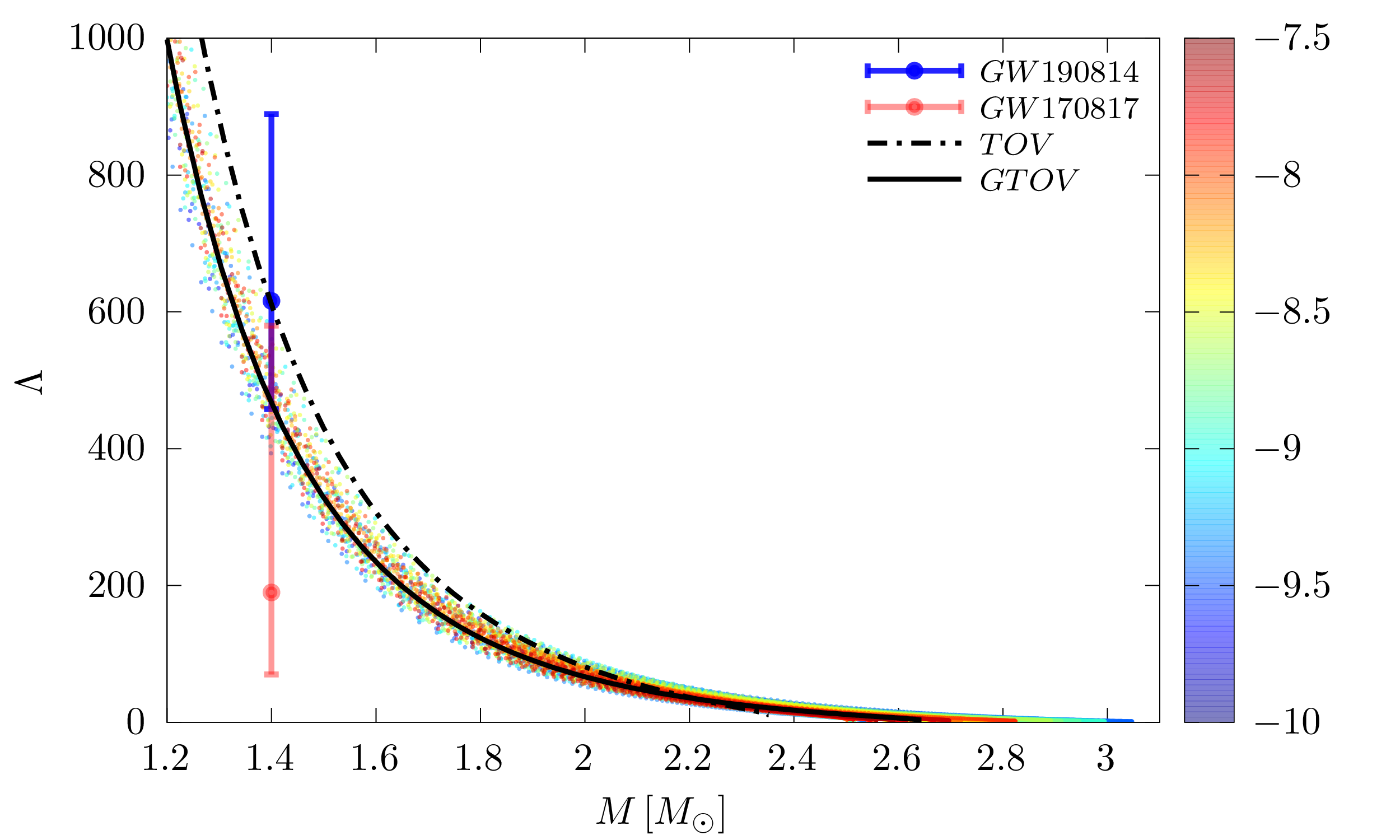}
    \caption{On top we show the corner plot of the posterior distribution of the GTOV parameters, on bottom left we show the mass-radius diagram and on bottom right we show the tidal deformability as a function of the mass. All plots are for Nl3$\omega\rho$Y EoS in the speculative scenario.} \label{nl3wpy_spec}
\end{figure}
%%%%%%%%%%%%%%%%%%%%%%%%%%%%%%%%%%%%%%%%%%%%%%%%%%%%%%%%%%%%%%%%%%%%%%%%%%%%%%%%%%%%

In this appendix we show the speculative scenario, were we consider the possibility that the compact object within the mass gap in event GW190814~\cite{LIGOScientific:2020zkf} is an NS, so we add its mass and tidal deformability as constraints to be fitted. This way, we look for values of deformability that satisfy the constraints from GW170817 and GW190814 at the same time, so that we also show the value of $\Lambda_{1.4}$ estimated for the GW190814 event, $\Lambda_{1.4} = 616^{+273}_{-158}$~\cite{LIGOScientific:2020zkf}, represented by a blue dot with an error bar in the plots for $\Lambda_{1.4}$. Additionally, in the figures for the speculative case also display the mass ($2.5^{+0.08}_{-0.09}$ M$_{\odot}$) of the possible NS in the event GW190814~\cite{LIGOScientific:2020zkf} as an orange continuous horizontal line and a shaded region for its the uncertainty. 

\subsection{El3$\omega\rho$}

For the El3$\omega\rho$ EoS, we observe that the GTOV parameters significantly deviate from those in General Relativity (GR). In particular, the parameter $\theta$ is negative in the speculative case, which contrasts with the positive value found in the conservative case. The mass-radius diagram for the speculative scenario shows that the GTOV models produce stars with relatively higher radii than the GR model, but these stars do not fully satisfy the NICER and GW170817 data. Additionally, in the speculative scenario, $\Lambda_{1.4}$ values fall within the lower margin of error for the GW190814 event. The speculative case generally predicts higher $\rm M_{max}$ and enhanced compactness for the stars compared to the GR case.

\subsection{Nl3$\omega\rho$}

For the Nl3$\omega\rho$ EoS, the GTOV parameters also deviate significantly from the GR results, with $\theta<0$ differing from the positive value in the conservative case. In GR this EoS model leads to a $\Lambda_{1.4}$ value that strongly satisfies GW190814. In the speculative scenario, the mass-radius data for GW190814 and NICER are satisfied but, this case slightly violates the GW170817 data with a smaller $\rm M_{max}$. The $\Lambda_{1.4}$ values fall in a range that satisfies both the GW170817 and GW190814 data. The mass-radius curves in this scenario deviate less from the GR case compared to the conservative case, and the stars tend to be less compact.

\subsection{El3$\omega\rho$Y}

In the El3$\omega\rho$Y model, once again the GTOV parameters show significant deviations relative to the GR model, and we obtain $\theta<0$. The inclusion of hyperons in the EoS reduces the maximum mass achievable in GR, failing to satisfy the PSR J0740$+$6620 and GW170817 mass constraints. On the other range, GTOV in the speculative case produces mass-radius relations that satisfy the NICER data and also align with the $\Lambda_{1.4}$ values that satisfy both the GW190814 and GW170817 events. However, the GTOV curves do not satify the GW170817 mass constraints. This shift in $\Lambda_{1.4}$ is toward the lower limit for the speculative case, while in the conservative case, it shifts toward the upper limit. The speculative scenario's GTOV models produce a significant increase in compactness, and the resulting stars align well with both GW events and NICER data.

\subsection{Nl3$\omega\rho$Y}

With Nl3$\omega\rho$Y EoS, the GTOV parameters also deviate significantly from the GR case, with $\theta$ negative in the speculative case and positive in the conservative one. Hyperonization leads to a reduction in $\rm M_{max}$ with relatively larger radii, which satisfies the PSR J0030$+$0451 mass-radius data in the GR case. In the speculative scenario, the GTOV models produce mass-radius relations that satisfy the GW190814 data, although they do not meet the GW170817 and PSR J0740$+$6620 constraints, due to the larger radii. 
\end{document}